\documentclass[fleqn,review,3p,times]{elsarticle}

\usepackage{amssymb}
\usepackage{amsthm}
\usepackage{adjustbox}
\usepackage{amsmath}
\usepackage{textcomp}
\usepackage{gensymb}
\usepackage{listings}
\usepackage{mathtools}
\usepackage{mathdots}
\usepackage{physics}
\usepackage[binary-units=true,exponent-product=\cdot]{siunitx}
\usepackage{subcaption}

\usepackage{shellesc}
\usepackage{tikz,tikzscale}
\usepackage{tikz-3dplot}
\usetikzlibrary{backgrounds}
\usetikzlibrary{calc}
\usetikzlibrary{fit}
\usetikzlibrary{intersections}
\usetikzlibrary{patterns}
\usetikzlibrary{through}

\tikzstyle{block} = [rectangle,draw,fill=gray!20,text width=5em,text centered, rounded corners, minimum height=4em,minimum width=20em,text width=20em]
\tikzstyle{cloud} = [diamond  ,draw,             text width=5em,text badly centered, node distance=3cm, inner sep=0pt]

\usepackage{pgfplots}
\pgfplotsset{compat=newest}
\pgfplotsset{lua backend=true}
\pgfdeclarelayer{background}
\pgfdeclarelayer{foreground}
\pgfsetlayers{background,main,foreground}

\makeatletter
\newcommand{\Distance}[3]{
\tikz@scan@one@point\pgfutil@firstofone($#1-#2$)\relax
\pgfmathsetmacro{#3}{round(0.99626*veclen(\the\pgf@x,\the\pgf@y)/0.0283465)/1000}}
\makeatother

\usepackage{tabularx}
\usepackage{multirow}
\newlength{\multiwidth}

\graphicspath{{fig/}}

\usepackage[hidelinks]{hyperref}
\usepackage{cleveref}

\lstdefinestyle{PICLas}{
  language=Fortran,
  basicstyle=\scriptsize\ttfamily\color{black!80},
  frame=tb,
  float,
  showspaces=false,
  showstringspaces=false,
  keywordstyle=\color{black},
  commentstyle=\color{darkgray},
  deletekeywords = {EQ},
  morekeywords={MPI_WIN_ALLOCATE_SHARED, MPI_WIN_SHARED_QUERY, C_F_POINTER, MPI_WIN_LOCK, MPI_WIN_SYNC, MPI_BARRIER},
  morecomment=[l]{!\ }
}

\mathchardef\mhyphen="2D

\newcommand\fnsep{\textsuperscript{,}}

\LetLtxMacro{\originaleqref}{\eqref}
\renewcommand{\eqref}{Eq.~\originaleqref}

\newcommand{\hawk}{\textit{hawk}}

\makeatletter
\def\ps@pprintTitle{%
 \let\@oddhead\@empty
 \let\@evenhead\@empty
 \def\@oddfoot{\centerline{\thepage}}%
 \let\@evenfoot\@oddfoot}
\makeatother

\begin{document}
\begin{frontmatter}

\title{Hybrid Parallelization of Euler-Lagrange Simulations Based on MPI-3 Shared Memory}

\author[ila]{Patrick Kopper\corref{cor1}}\cortext[cor1]{Corresponding author.}\ead{kopper@ila.uni-stuttgart.de}
\author[boltzplatz]{Stephen Copplestone}\ead{copplestone@boltzplatz.eu}
\author[irs]{Marcel Pfeiffer}\ead{mpfeiffer@irs.uni-stuttgart.de}
\author[ila]{Christian Koch}\ead{christian.koch@ila.uni-stuttgart.de}
\author[irs]{Stefanos Fasoulas}\ead{fasoulas@irs.uni-stuttgart.de}
\author[isut]{Andrea Beck}\ead{andrea.beck@ovgu.de}
\affiliation[ila]{organization={University of Stuttgart, Institute of Aircraft Propulsion Systems},
            addressline={Pfaffenwaldring 6},
            city={Stuttgart},
            country={Germany}}

\affiliation[boltzplatz]{organization={boltzplatz - numerical plasma dynamics GmbH},
            addressline={Schelmenwasenstr. 34},
            city={Stuttgart},
            country={Germany}}

\affiliation[irs]{organization={University of Stuttgart, Institute of Space Systems},
            addressline={Pfaffenwaldring 29},
            city={Stuttgart},
            country={Germany}}

            \affiliation[isut]{organization={Otto-von-Guericke University Magdeburg, Laboratory of Fluid Dynamics and Technical Flows},
            addressline={Universitaetsplatz 2},
            city={Magdeburg},
            country={Germany}}

\begin{abstract}
The use of Euler-Lagrange methods on unstructured grids extends their application area to more versatile setups. However, the lack of a regular topology limits the scalability of distributed parallel methods, especially for routines that perform a physical search in space. One of the most prominent slowdowns is the search for halo elements in physical space for the purpose of runtime communication avoidance. In this work, we present a new communication-free halo element search algorithm utilizing the MPI-3 shared memory model. This novel method eliminates the severe performance bottleneck of many-to-many communication during initialization compared to the distributed parallelization approach and extends the possible applications beyond those achievable with the previous approach. Building on these data structures, we then present methods for efficient particle emission, scalable deposition schemes for particle-field coupling, and latency hiding approaches. The scaling performance of the proposed algorithms is validated through plasma dynamics simulations of an open-source framework on a massively parallel system, demonstrating an efficiency of up to $80\si{\percent}$ on \num{131000} cores.
\end{abstract}


\end{frontmatter}

\section{Introduction}
\label{sec:introduction}
The initialization time, i.e., the time from beginning of the code execution until the first computation step, plays a critical role in Euler-Lagrangian solvers in a high-performance computing context as it is closely linked with adequate load balancing. Ideally, each processor should receive equal load to achieve maximum overall simulation efficiency. Accurate load estimation is already challenging in the case of a pure Euler solver as various cell sizes, time steps, local models and boundary conditions must be considered. Nonetheless, there are effective techniques to determine the local load and thus the grid distribution a priori, also known as static load balancing~\cite{Tantawi1985,Kameda1997,Sukys2012}. For these cases, a prolonged initialization period is acceptable if it results in improved runtime performance.

The presence of the Lagrangian phase adds substantial complexity as discrete particles introduce additional load, which is only weakly correlated with the local element sizes. Furthermore, particle concentrations may shift during the simulation, with high fluid and particle loads often occurring at the same mesh location, especially in fluid simulations. Load balancing approaches in Euler-Lagrangian solvers must adapt to these changes during runtime, which is referred to as dynamic load balancing. Over time, various load distribution strategies have evolved, which can generally be classified into two categories: 1) task parallelization and 2) domain partitioning. Task parallelization splits the work along the phase interface, distributing the fluid work and the particle work to different processors. The advantage is that due to the same nature of work within a phase, both groups of processors can internally subdivide the overall task in an optimal way. Implementations of this approach have been presented e.g., in Refs.~\cite{Frank2001,Vance2002,Darmana2006}. The downside of task parallelization is the loss of any local connectivity, resulting in large communication effort, great memory requirements or both, rendering it inadequate for massively parallel computation where memory and interconnect bandwidth are scant~\cite{Kormann2011}. Domain decomposition keeps the locality between the two phases intact but requires the load distribution to be performed on the combined work with examples of this approach published in Refs.~\cite{Liewer1989,Rossi2013,Mehrling2014,Surmin2016}.  Additionally, communication patterns become unpredictable as a processor may receive elements of discrete phase but not necessarily transmit them and vice versa~\cite{Carmona1997}. Nevertheless, most massively parallel codes use domain decomposition, an approach we also follow.

As the focus of this work is on the discrete phase, an efficient solver for the continuous phase is presumed. Modern CFD solutions require high scalar performance and preferably minimal communication as storage and communication resources are unable to keep up with the steadily increase of available computing power~\cite{Hennessy2017,Roser2018}. High-order codes based on the Discontinuous Galerkin Spectral Element Method (DGSEM) have emerged as a well-suited approach as the fluid phase requires only the exchange of flux information at an element face, leading to a highly efficient numerical scheme while at the same time having dense, local operations~\cite{Kopriva2009}. Implementation as a solver for unstructured grids with possibly curved elements facilitates the creation of body-fitted domains even for complex geometries while retaining the high-order accuracy~\cite{Bassi1997}. The presence of the Lagrangian phase necessitates dynamic load balancing, which should be regularly performed as particle loads can heavily shift during the simulation. Hereby, time spent on the load balancing step has to be kept minimal in order not to counteract benefits in application performance~\cite{Garcia2014}. However, this task is non-trivial. While the DGSEM leads to a highly local scheme for the continuous phase, following the distributed memory approach means that each processor contains only local information on both the solution, associated quantities and the mesh without ready access to adjacent grid information in the case of distributed I/O. The absence of this mesh information prevents full tracking of a particle in the event of it crossing a partition boundary if not remedied.

One method that allows the tracking of particles across different partitions is based on the idea of a shared layer of elements surrounding each partition, an idea akin to the ghost cell approach in e.g., finite volume methods. These halo regions contain the geometric information of neighboring cells within a given distance, here referred to as the halo distance, from the local domain and enable the completion of particle tracking on the initial processor, thus delaying the need for communication and requiring only the exchange of particles after having crossed the domain boundary~\cite{Fasoulas2019}. While the process of halo element identification is straightforward for structured grids, it becomes significantly more complex for unstructured approaches as given within the present framework. Here, the search must be performed in physical space as the grid cells within a spatial region are not trivially mapped to locations in the mesh file~\cite{Ortwein2019}.

Performing the search exclusively on processor-local mesh information, i.e., an inward search from the processor MPI border, leads to a severe performance bottleneck as the grid information for each cell residing on a single processor needs to be sent to a multitude of other processors within the halo distance, requiring many-to-many or in the worst case all-to-all point communication. This congests the interconnect infrastructure with the potential to stall code execution for minutes or even hours. As the information which needs to be exchanged grows both with grid size and number of processors involved in the simulation, the issue only becomes more urgent on modern massively parallel architectures where memory and communication constraints are all the more apparent. Moreover, as runtime load balancing is generally desirable to counter the shifting of particle loads during the simulation, the identification of the halo region must be performed multiple times during a given simulation, thus prompting the need for an efficient scheme which avoids detrimental effects on overall simulation performance. While this constitutes a new challenge for unstructured approaches that is absent in structured grids, this is easily outweighed by the advantages of such unstructured approaches for body fitted grids in domains of practical relevance.

Towards this goal, we present in this work a novel approach to unstructured Euler-Lagrangian simulations based on MPI-3 shared memory. Within this approach, we store information with compute node granularity and perform a multi-step communication-free parallel search on the compute node to identify the elements in the halo region, thereby retaining excellent scaling properties on today's massively parallel supercomputing architectures. Building on these data structures, we present methods for efficient particle emission, scalable deposition schemes for particle-field coupling, and latency hiding approaches. These efforts then give us the ability to conduct high-order simulations of rarefied gas flows at an industrial scale beyond generic test cases. To the best of our knowledge, this is the first unstructured framework enabling massively-parallel Euler-Lagrangian simulations at this problem size. The implementation considered in the present work is open-source and available on GitHub\footnote{\url{https://github.com/flexi-framework/flexi}}\fnsep\footnote{\url{https://github.com/piclas-framework/piclas}}.

The outline of this paper is as follows:
The governing equations for non-equilibrium gas flows followed by the DGSEM scheme as well as the theory for particle motion and tracking are given in~\cref{sec:theory}. A high-level overview on the parallelization strategy is given in~\cref{sec:strategy}. In~\cref{sec:implementation}, we present the shared-memory approach for halo region determination and distribution, thereby shifting the communication load from the processor to the compute node level and alleviating the aforementioned scaling restrictions. The remainder of this section is designated to methods for emission, deposition on latency hiding based on the same shared-memory approach. The test case of an adiabatic box representing the optimum for the parallelization concept as well as near-application cases of a supersonic flow and a gyrotron resonator are discussed in~\cref{sec:testcases}, and the scaling results are presented in section~\cref{sec:results}. We conclude with a brief summary and give an outlook on further developments in~\cref{sec:conclusions}.

\section{Theory}
\label{sec:theory}
The approaches presented here are applicable to any unstructured Euler-Lagrange code and are already implemented in the two high-order open-source frameworks FLEXI\footnote{\url{https://www.flexi-project.org}}~\cite{Hindenlang2012,Krais2021} and PICLas~\cite{Fasoulas2019,Munz2014}. Both are actively developed at University of Stuttgart and share a common code basis for the DGSEM solver with FLEXI solving a continuous fluid phase prescribed by the compressible Navier-Stokes-Fourier equations while PICLas uses Maxwell's equations for the electromagnetic fields. FLEXI currently focuses on inertial particles in turbomachinery applications~\cite{Beck2019} with initial performance for the current methods presented in~\cite{Kopper2021}. The present work focuses on solutions to non-equilibrium gas and plasma flows within the PICLas framework using the Particle-In-Cell (PIC) approach~\cite{birdsall199101,Hockney1988} as well as a particle based Bhatnagar-Gross-Krook (BGK) solver~\cite{bhatnagar1954model,pfeiffer2018particle}.

\subsection{Non-equilibrium gas flows}
\label{sec:theory:Plasma}
Non-equilibrium gas and plasma flows are generally characterized by possibly charged particles that interact with an electromagnetic field where the statistical distribution is described by Boltzmann's equation~\cite{Bird1994,birdsall199101}
\begin{equation}
  \frac{\partial f}{\partial t} + \mathbf{v}\frac{\partial f}{\partial\mathbf{x}} + \frac{\mathbf{F}}{m}\frac{\partial f}{\partial\mathbf{v}} = \left.\frac{\partial f}{\partial t}\right\vert_\text{coll}.
  \label{eqn:boltzmann}
\end{equation}
Here, $f = f(\mathbf{x},\mathbf{v},t)$ represents the probability distribution function, i.e., the expected particle density at position $\mathbf{x}$ with velocity $\mathbf{v}$. In the methods used here, the distribution function is approximated by particles. These can move freely and represent the phase space from a Lagrangian point of view through their location and velocity. The particles interact with each other through the right hand side described by a collision operator whereas the interaction for charged particles occurs through the Lorenz force $\mathbf{F}=\mathbf{F}_L$. The Lorentz forces are calculated from electromagnetic fields, which are solved on a fixed grid in an Eulerian fashion. Typically, the Boltzmann collision integral is used as collision operator which effectively gives the change of the particle probability density function caused by binary particle-particle collisions.

However, since the Boltzmann collision integral is numerically difficult and time-consuming for various reasons~\cite{Bird1994}, it is often approximated, e.g., by the Fokker–Planck (FP) solution algorithm~\cite{Jenny2010,Gorji2011} or the BGK approximation~\cite{bhatnagar1954model,pfeiffer2018particle, zhang2019particle}, see~\cref{sec:theory:BGK}.
Beyond particle-particle collisions, charged particles experience the Lorentz force
\begin{equation}
  \mathbf{F}_L = q \left(\mathbf{E} + \mathbf{v} \cross \mathbf{B} \right)
  \label{eqn:lorentz}
\end{equation}
Here, $q$ is the electric charge of a given particle whereas the electric field $\mathbf{E}$ and magnetic field $\mathbf{B}$ obey Maxwell's equations~\cite{Jackson1999_3rd}
\begin{align}
  \frac{\partial \mathbf{D}}{\partial t} &= \nabla \cross \mathbf{H} - \mathbf{j}\\
  \frac{\partial \mathbf{B}}{\partial t} &= -\nabla \cross \mathbf{E}\\
  \nabla \cdot \mathbf{D}                &= \rho\\
  \nabla \cdot \mathbf{B}                &= 0
  \label{eqn:Maxwell}
\end{align}
with $\mathbf{D}$ being the electric displacement field, $\mathbf{H}$ the magnetic field strength whereas $\rho$ and $\mathbf{j}$ represent the charge and current density, respectively. The field equations from \eqref{eqn:Maxwell} are thereby solved on a fixed grid, i.e. the Eulerian view. The coupling between the Eulerian view and the Lagrangian view arises on the one hand through the Lorentz force \eqref{eqn:lorentz}. Here, the fields are interpolated from the solution on the Euler grid to the Lagrangian particles in order to calculate the forces acting on these particles. On the other hand, the charge densities and current densities as source terms of the Maxwell equations on the Euler grid correspond to the zeroth and first moment of the distribution function, i.e., they are obtained by interpolating the particle data to the fixed Euler grid.

\subsection{Discontinuous Galerkin Spectral Element Method (DGSEM)}
\label{sec:theory:DGSEM}
In order to enforce charge conservation, Maxwell's equations are cast into the purely hyperbolic Maxwell (PHM) form~\cite{munz2000divergence} which can then be solved using the high-order Discontinuous Galerkin Spectral Element Method (DGSEM)~\cite{Sonntag2017,kopriva2002computation}. DG methods operate on a weak formulation of the conservation equations following the method of lines approach by projecting them onto a space of polynomial test functions in reference space. Collocation of interpolation and integration points yields a highly efficient scheme which is advanced by an explicit Runge-Kutta scheme in time. PICLas is designed as a solver for unstructured grids, thereby allowing the straightforward creation of body-fitted grids with curved boundaries even for complex geometries. The approximation of the solution by a high order polynomial in each element ensures that high-order accuracy is retained~\cite{Bassi1997}.

\subsection{Particle Behavior}
\label{sec:theory:particles}
Particles contribute to the electrical and magnetic field (deposition, see~\cref{sec:implementation:shape}) but are simultaneously influenced by the field through the Lorentz force and optionally through particle-particle collisions.

\subsubsection{Equation of Motion}
Based on~\cref{eqn:lorentz}, the change in position and velocity of each particle is given by the relativistic equation of motion,
\begin{align}
  \frac{d\mathbf{x}_p}{dt}       &= \mathbf{v}_p\label{eqn:particle:motion1}\\
  \frac{d\gamma\mathbf{v}_p}{dt} &= \frac{q_p}{m_p}\left(\mathbf{E} + \mathbf{v}\times\mathbf{B}\right),\label{eqn:particle:motion2}
\end{align}
with $\gamma$ being the Lorentz factor given as
\begin{equation}
  \gamma = \left(1-\frac{\abs{\mathbf{v}}^2}{c^2}\right)^{-1/2}.
\end{equation}
Here, $\mathbf{x}$ and $\mathbf{v}$ represent the position and velocity of a given particle in physical space, $q$ and $m$ its charge and mass, and $c$ the speed of light.

\subsubsection{Bhatnagar-Gross-Krook (BGK)}
\label{sec:theory:BGK}
The BGK operator approximates the collision term in \eqref{eqn:boltzmann} to a simple relaxation form where the distribution function relaxes towards a target distribution function $f^t$ with a
certain relaxation frequency $\nu$:
\begin{equation}
\left.\frac{\partial f}{\partial t}\right|_{Coll}=\nu\left(f^t-f\right).
\label{eq:bgkmain}
\end{equation}
The original BGK model assumes that the target velocity distribution function is the Maxwellian velocity distribution
\begin{equation}
f^M=n\left(\frac{m}{2\pi k_B T}\right)^{3/2} \exp\left[-\frac{m\mathbf c^2}{2k_B T}\right],
\label{eq:maxwelldist}
\end{equation}
with the particle density $n$, particle mass $m$, temperature $T$ and the thermal particle velocity $\mathbf c=\mathbf v -\mathbf u$ from the
particle velocity $\mathbf v$ and the average flow velocity $\mathbf u$~\cite{bhatnagar1954model}. In order to obtain the correct Prandtl number in the flow, more complex target distribution functions must be used. A very frequently used target distribution function that is also applied here is the ellipsoidal statistical BGK target function, see details in~\citet{holway1966new}.
In PICLas, the solution of the BGK equation is performed in a pure Lagrangian and stochastic manner with particles using the stochastic particle Bhatnagar-Gross-Krook (SP-BGK) method as described in \cite{pfeiffer2018extending,pfeiffer2019extension}.
It offers an efficient approach to simulate non-equilibrium flows in smaller Knudsen number regimes. All particles within a cell interact with each other by the relaxation process described in \eqref{eq:bgkmain}. This is similar to the collision process in the well-known Direct Simulation Monte Carlo (DSMC) method~\cite{Bird1994}, which also happens only within a cell. Subsequently, the convection due to the free movement of the particles is modeled together with boundary conditions in the computational domain in order to perform a relaxation process again.

\subsection{Localization and Tracking}
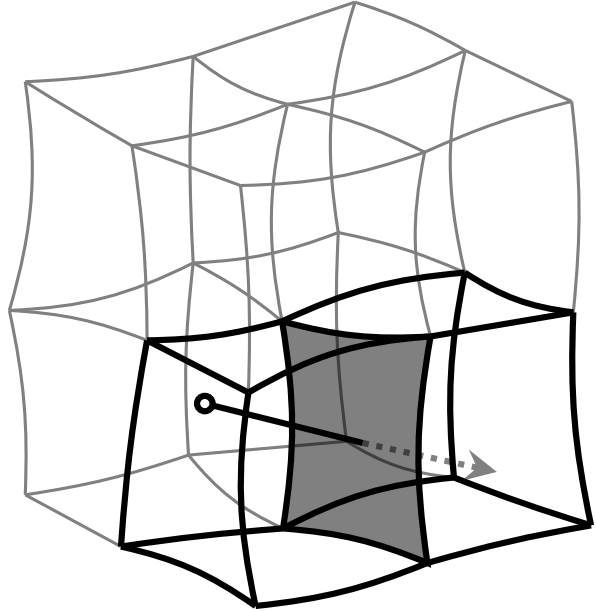
\begin{figure}[htb!]
\centering\begin{adjustbox}{width=.5\textwidth}
\def\NodeSize{6pt}

\begin{tikzpicture}[x={(0.535cm,-0.191cm)},y={(0cm,0.943cm)},z={(-0.754cm,-0.182cm)}]
  \pgfmathsetseed{6}%
  \def\randfac{10}
  \def\randbend{15}

  \coordinate(A1) at ( 0.0+rand/\randfac,  0.0+rand/\randfac,  0.0+rand/\randfac);
  \coordinate(B1) at ( 1.0+rand/\randfac,  0.0+rand/\randfac,  0.0+rand/\randfac);
  \coordinate(C1) at ( 0.0+rand/\randfac,  1.0+rand/\randfac,  0.0+rand/\randfac);
  \coordinate(D1) at ( 1.0+rand/\randfac,  1.0+rand/\randfac,  0.0+rand/\randfac);
  \coordinate(E1) at ( 0.0+rand/\randfac,  0.0+rand/\randfac,  1.0+rand/\randfac);
  \coordinate(F1) at ( 1.0+rand/\randfac,  0.0+rand/\randfac,  1.0+rand/\randfac);
  \coordinate(G1) at ( 0.0+rand/\randfac,  1.0+rand/\randfac,  1.0+rand/\randfac);
  \coordinate(H1) at ( 1.0+rand/\randfac,  1.0+rand/\randfac,  1.0+rand/\randfac);

  \pgfmathsetseed{9}%

  \begin{pgfonlayer}{foreground}
  \draw[bend right=rand*\randbend,thick] (E1) to (F1);
  \draw[bend right=rand*\randbend,thick] (E1) to (G1);
  \draw[bend right=rand*\randbend,thick] (F1) to (H1);
  \draw[bend right=rand*\randbend,thick] (G1) to (H1);

  \draw[bend right=rand*\randbend,thick] (A1) to (E1);
  \draw[bend right=rand*\randbend,thick] (B1) to (F1);
  \draw[bend right=rand*\randbend,thick] (C1) to (G1);
  \draw[bend right=rand*\randbend,thick] (D1) to (H1);
  \end{pgfonlayer}

  \begin{pgfonlayer}{main}
  \draw[bend left =0.6*\randbend,thick,fill=black, fill opacity=.5] (A1) to (B1) to (D1) to (C1) to (A1);
  \end{pgfonlayer}

  \coordinate(A2) at ( 0.0+rand/\randfac,  0.0+rand/\randfac, -1.0+rand/\randfac);
  \coordinate(B2) at ( 1.0+rand/\randfac,  0.0+rand/\randfac, -1.0+rand/\randfac);
  \coordinate(C2) at ( 0.0+rand/\randfac,  1.0+rand/\randfac, -1.0+rand/\randfac);
  \coordinate(D2) at ( 1.0+rand/\randfac,  1.0+rand/\randfac, -1.0+rand/\randfac);

  \pgfmathsetseed{9}%
  \begin{pgfonlayer}{foreground}
  \draw[bend right=rand*\randbend,thick] (A1) to (A2);
  \draw[bend right=rand*\randbend,thick] (B1) to (B2);
  \draw[bend right=rand*\randbend,thick] (C1) to (C2);
  \draw[bend right=rand*\randbend,thick] (D1) to (D2);

  \draw[bend right=rand*\randbend,thick] (A2) to (B2);
  \draw[bend right=rand*\randbend,thick] (A2) to (C2);
  \draw[bend right=rand*\randbend,thick] (B2) to (D2);
  \draw[bend right=rand*\randbend,thick] (C2) to (D2);
  \end{pgfonlayer}

  \coordinate(C3) at ( 0.0+rand/\randfac,  2.0+rand/\randfac, 0.0+rand/\randfac);
  \coordinate(D3) at ( 1.0+rand/\randfac,  2.0+rand/\randfac, 0.0+rand/\randfac);
  \coordinate(G3) at ( 0.0+rand/\randfac,  2.0+rand/\randfac, 1.0+rand/\randfac);
  \coordinate(H3) at ( 1.0+rand/\randfac,  2.0+rand/\randfac, 1.0+rand/\randfac);

  \pgfmathsetseed{6}%
  \draw[bend right=rand*\randbend,gray] (C1) to (C3);
  \draw[bend right=rand*\randbend,gray] (D1) to (D3);
  \draw[bend right=rand*\randbend,gray] (G1) to (G3);
  \draw[bend right=rand*\randbend,gray] (H1) to (H3);

  \draw[bend right=rand*\randbend,gray] (C3) to (D3);
  \draw[bend right=rand*\randbend,gray] (C3) to (G3);
  \draw[bend right=rand*\randbend,gray] (G3) to (H3);
  \draw[bend right=rand*\randbend,gray] (D3) to (H3);

  \coordinate(C4) at ( 0.0+rand/\randfac,  2.0+rand/\randfac, -1.0+rand/\randfac);
  \coordinate(D4) at ( 1.0+rand/\randfac,  2.0+rand/\randfac, -1.0+rand/\randfac);

  \pgfmathsetseed{3}%
  \draw[bend right=rand*\randbend,gray] (C2) to (C4);
  \draw[bend right=rand*\randbend,gray] (D2) to (D4);
  \draw[bend right=rand*\randbend,gray] (C4) to (D4);
  \draw[bend right=rand*\randbend,gray] (C3) to (C4);
  \draw[bend right=rand*\randbend,gray] (D3) to (D4);

  \coordinate(A5) at (-1.0+rand/\randfac,  0.0+rand/\randfac, 0.0+rand/\randfac);
  \coordinate(C5) at (-1.0+rand/\randfac,  1.0+rand/\randfac, 0.0+rand/\randfac);
  \coordinate(E5) at (-1.0+rand/\randfac,  0.0+rand/\randfac, 1.0+rand/\randfac);
  \coordinate(G5) at (-1.0+rand/\randfac,  1.0+rand/\randfac, 1.0+rand/\randfac);

  \pgfmathsetseed{3}%
  \draw[bend right=rand*\randbend,gray] (A1) to (A5);
  \draw[bend right=rand*\randbend,gray] (C1) to (C5);
  \draw[bend right=rand*\randbend,gray] (E1) to (E5);
  \draw[bend right=rand*\randbend,gray] (G1) to (G5);

  \draw[bend right=rand*\randbend,gray] (A5) to (C5);
  \draw[bend right=rand*\randbend,gray] (A5) to (E5);
  \draw[bend right=rand*\randbend,gray] (E5) to (G5);
  \draw[bend right=rand*\randbend,gray] (C5) to (G5);

  \coordinate(C6) at (-1.0+rand/\randfac,  2.0+rand/\randfac, 0.0+rand/\randfac);
  \coordinate(G6) at (-1.0+rand/\randfac,  2.0+rand/\randfac, 1.0+rand/\randfac);

  \pgfmathsetseed{4}%
  \draw[bend right=rand*\randbend,gray] (C5) to (C6);
  \draw[bend right=rand*\randbend,gray] (G5) to (G6);
  \draw[bend right=rand*\randbend,gray] (C3) to (C6);
  \draw[bend right=rand*\randbend,gray] (G3) to (G6);
  \draw[bend right=rand*\randbend,gray] (C6) to (G6);

  \coordinate(A7) at (-1.0+rand/\randfac,  0.0+rand/\randfac,-1.0+rand/\randfac);
  \coordinate(C7) at (-1.0+rand/\randfac,  1.0+rand/\randfac,-1.0+rand/\randfac);

  \pgfmathsetseed{3}%
  \begin{pgfonlayer}{background}
  \draw[bend right=rand*\randbend,gray] (A2) to (A7);
  \draw[bend right=rand*\randbend,gray] (C2) to (C7);
  \draw[bend right=rand*\randbend,gray] (A5) to (A7);
  \draw[bend right=rand*\randbend,gray] (C5) to (C7);
  \draw[bend right=rand*\randbend,gray] (A7) to (C7);
  \end{pgfonlayer}

  \coordinate(C8) at (-1.0+rand/\randfac,  2.0+rand/\randfac,-1.0+rand/\randfac);

  \pgfmathsetseed{3}%
  \begin{pgfonlayer}{background}
  \draw[bend right=rand*\randbend,gray] (C7) to (C8);
  \draw[bend right=rand*\randbend,gray] (C4) to (C8);
  \draw[bend right=rand*\randbend,gray] (C6) to (C8);
  \end{pgfonlayer}

  \begin{pgfonlayer}{main}
  \draw[thick,black]                            (0.2, 0.7, 0.7) -> (0.55, 0.445, 0);
  \draw[thick,black,fill=white] (0.2, 0.7, 0.7) circle[radius=1pt];
  \end{pgfonlayer}
  \begin{pgfonlayer}{background}
  \draw[thick,black!50,->,>=stealth,densely dotted]        (0.55, 0.445, 0) -> (0.7, 0.2, -0.7);
  \end{pgfonlayer}
\end{tikzpicture}
\end{adjustbox}
\caption{Particle tracing in physical space}
\label{fig:particle:tracing}
\end{figure}

Particles are tracked in physical space for all cases considered within this work. The first step within this approach is the solution of~\cref{eqn:particle:motion1,eqn:particle:motion2} to obtain the new particle position. In order to identify a particle crossing an element boundary and its recipient, all faces of the previous element are checked for intersections with the particle path, see~\cref{fig:particle:tracing}. This procedure is performed iteratively until no more intersections are found and thus the final element is determined. If an element face is representing a boundary face, the corresponding boundary conditions can intuitively be incorporated by adjusting the remaining particle path. For more details and an alternative tracking approach based on localization in the reference space, see~\cite{Ortwein2019}.

\section{Parallelization Strategy}
\label{sec:strategy}
High-performance clusters are almost exclusively constructed as distributed systems, connecting separate nodes through an interconnect. In general, any interconnect imposes bandwidth starvation compared to local memory while simultaneously incurring latency costs. Thus, efficient parallelization approaches need to employ two strategies: 1) Communication avoidance and 2) latency hiding. By reducing the amount of transferred data, congestion on the interconnect can be alleviated. Performing the communication in a non-blocking manner allows local work to continue, thereby obfuscating the additional latency of the interconnect.

\subsection{Continuous Phase}
Relying on the DGSEM allows PICLas to make extensive use of both strategies. By enforcing a basis with local support, the volume integral becomes a purely local operation and only the surface flux information has to be exchanged on the cell boundaries. To ensure fast initialization times, the unstructured fluid elements are pre-sorted along a space-filling curve (SFC) during mesh generation. SFC have shown their suitability for efficient calculation of new distributions during runtime for load balancing purposes in PIC simulations~\cite{Harlacher2012,Germaschewski2016}. Furthermore, the SFC allows for highly parallel, non-overlapping disk storage access with an arbitrary number of processors~\cite{Atak2016}. Details on the implementation are given in~\cite{Krais2021}.

\subsection{Halo Region}
Halo regions bring these two strategies from continuous Eulerian to discrete Lagrangian phase. Since particle tracking is performed in unstructured physical space, geometric information along the considered particle path must be available at the time of tracking. By enriching the local DG domain with geometric information up to given physical distance from the domain boundaries, each processor can complete tracking to the final particle position. This halo distance is chosen as the maximum possible distance any particle can travel within a simulation time step, thereby ensuring a processor has all eligible elements accessible while simultaneously generating the minimal number of halo elements. Using this approach, communication is delayed until a particle is found to have left the processor domain after accounting for boundary conditions and only the minimum required information, i.e., the particle properties including the new particle position, must be communicated.

However, this shifts some work from the simulation time-stepping to the routines where geometric and neighboring information must be established or updated. This corresponds to the initialization and any load balancing step, which always also includes an update of this information. PICLas follows the commonly used restart-based load-balancing approach where the simulation is saved to disk and reloaded using an improved load distribution. As a result, this procedure relies on fast initialization times and is aided by the approaches outlined in the previous section. However, given the requirement to work on unstructured meshes, the required halo elements for the particulate phase can only be determined through a search in physical space even with the mesh elements already pre-sorted along the SFC. As was previously outlined, performing the search and subsequent communication of mesh information using only processor-local information incurs severe performance penalties stemming from the differences in processor work from load distribution and the required many-to-many communication. The latter case in particular is exacerbated by modern many-core architectures, which consequently limits the scaling of the approach and necessitates the novel approach outlined in the subsequent~\cref{sec:implementation}.

\subsection{Load Balancing}
\label{sec:theory:loadbalancing}
Immediate benefit of the improved performance for the restart-based load balancing is the ability to increase the number of load evaluation and - if necessary - balancing steps. Following the classification by Watts and Taylor~\cite{Watts1998}, the load evaluation can be based on the application, i.e., a priori using information on the algorithms involved; the system, i.e., at runtime using timing information, or a combination of both. While the application-based approach is simple to implement and successfully used for single-phase simulations with constant computation and communication time per degree of freedom~\cite{Krais2021}, the determination of the correct weights becomes challenging for multi-phase flows. Hence, a hybrid approach is commonly considered more robust.\\\indent
The approach utilized by PICLas relies on runtime measurements of the field solver and particle solver. Load evaluation steps are performed by comparing the runtime per rank at user-defined intervals. For this, total time spent in the field and particle solver are recorded prior to the load balancing step. Particles are assumed to remain in their element sufficiently long to assign their load to the element they are currently residing in. Thus, the load of a given element is estimated by the combination of the total time spent in field and particle solver, divided by the local number of elements and the element's contribution to the total tracking steps, respectively. The resulting time for a respective element is then given as
\begin{equation}
  t_\text{elem,tot} = t_\text{field} + \delta t_\text{particle} \cdot n_\text{particles,elem}.
\end{equation}
If an imbalance exceeding an acceptable threshold is detected, each rank gets assigned a new range of elements such that the load deviation becomes minimal, i.e.,
\begin{equation}
  \Delta t_\text{tot} = \abs{t_\text{target} - \sum_{\text{i}_\text{start}}^{\text{i}_\text{end}} t_\text{i,tot}} \stackrel{!}{=} min
\end{equation}
where $\text{i}_\text{start}$ and $\text{i}_\text{end}$ correspond to the respective element indices along the space-filling curve. More details on this implementation can be found in~\cite{Ortwein2018}.

\section{Implementation}
\label{sec:implementation}
This section describes the parallel implementation of the halo region search, emission and runtime deposition mechanisms. Following the MPI-3 shared memory paradigm, we store information with compute node granularity and perform subsequent routines in a communication-free way on the shared memory region. This chapter serves to illustrate the allocation of the shared memory window, the mesh distribution and aforementioned routines in~\cref{sec:implementation:allocation},~\cref{sec:implementation:mesh} and~\cref{sec:implementation:halo,sec:implementation:emission,sec:implementation:shape}, respectively. Where applicable, we print the Fortran source code rather than pseudo-code to facilitate implementation in other scientific frameworks. Further information on how the stored mesh information is subsequently used for particle tracking is given in~\cite{Ortwein2019}.

\subsection{Shared Memory Allocation}
\label{sec:implementation:allocation}
Classical shared memory programming involves OpenMP. However, this approach is limited to single-node cases as OpenMP cannot handle distributed memory. Message Passing Interface (MPI)-3 introduces the concept of shared memory with the MPI Shared Memory (SHM) model. The resulting coding approach is also called ''hybrid parallel programming'' as it combines the shared memory approach of OpenMP with the distributed memory view of MPI. Memory regions allocated with MPI-3 SHM can be distributed arbitrarily between the processors while being accessible by all processors on the compute node. In our implementation for cache-coherent systems (\verb!MPI_WIN_UNIFIED!), the shared memory window is allocated only on the compute node root to avoid offset calculations. The shared window is continuous in memory and thus can be directly read by each process. We do not employ RMA routines for store operations but ensure non-overlapping writes through data distribution along the SFC. \verb!MPI_WIN_SYNC! calls ensure explicit synchronization and immediate availability of the written information on the compute node. The code for this approach is given in~\cref{lst:impl:mpi3}.

\begin{lstlisting}[style=PICLas,caption=Example storage of ELEMINFO array in shared memory,label={lst:impl:mpi3}]
! Only compute node MPI root actually allocates the memory
WIN_SIZE = MERGE(DATASIZE_BYTE,INT(0,MPI_ADDRESS_KIND),MYCOMPUTENODERANK.EQ.0)
CALL MPI_WIN_ALLOCATE_SHARED(WIN_SIZE,DISP_UNIT,MPI_INFO,MPI_COMM_SHARED,SHM_PTR,SHM_WIN,IERROR)

! Node MPI root already knows the location in virtual memory, all other find it here
IF (MYCOMPUTENODERANK.NE.0) CALL MPI_WIN_SHARED_QUERY(SHM_WIN,0,WIN_SIZE,DISP_UNIT,SHM_PTR,IERROR)

! SHM_PTR can now be associated with a Fortran pointer and thus used to access the shared data
CALL C_F_POINTER(SHM_PTR,DATAPOINTER,NVAL)

! Start passive RMA epoch
CALL MPI_WIN_LOCK_ALL(INFO,SHM_WIN,IERROR)

! Copy element info into shared memory
ELEMINFO_SHARED(1:ELEMINFOSIZE,OFFSETELEM+1:OFFSETELEM+NELEMS) = ELEMINFO(1:ELEMINFOSIZE,1:NELEMS)

! Synchronize public and private copies
CALL MPI_WIN_SYNC(SHM_WIN,IERROR)
CALL MPI_BARRIER(MPI_COMM_SHARED,iError)
\end{lstlisting}


\subsection{Mesh Distribution}
\label{sec:implementation:mesh}
\begin{figure}[htb!]
\begin{subfigure}[t]{\textwidth}
\vspace{0pt}
\begin{adjustbox}{width=\textwidth}
\def\NodeSize{6pt}
\input{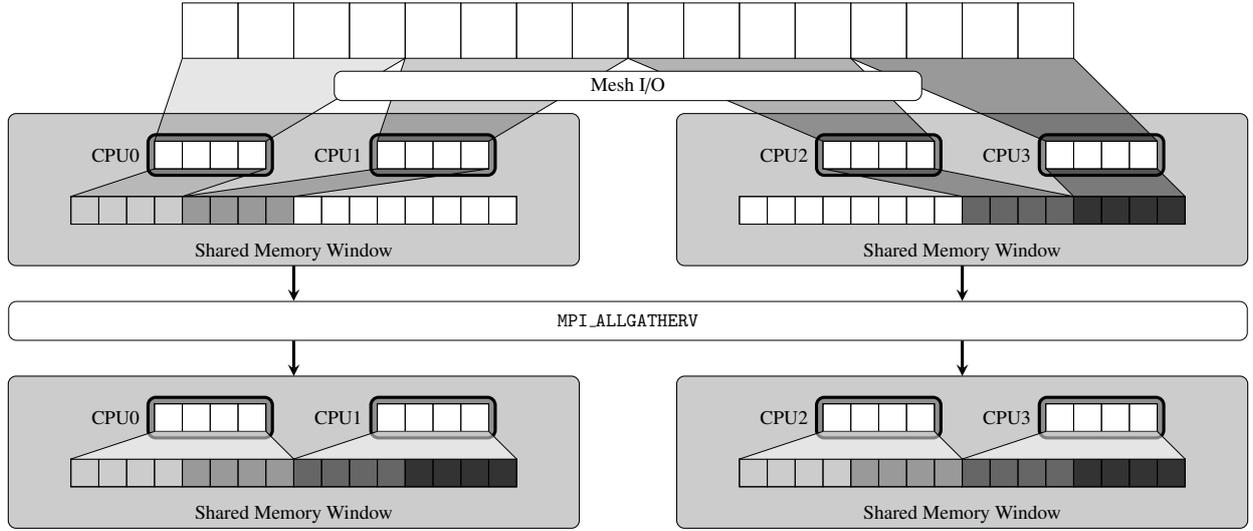}
\end{adjustbox}
\end{subfigure}
\caption{Mesh I/O along space-filling curve and distribution in shared memory. Additionally, the bottom part shows the processor assignments for the first step of the halo element search algorithm.}
\label{fig:impl:halo:readin}
\end{figure}

High-order meshes are created with the in-house preprocessor \textit{HOPR}\footnote{\url{https://github.com/hopr-framework/hopr}}~\cite{HOPR}. The mesh elements are ordered along a space-filling curve and saved in binary HDF5 format for highly parallel access~\cite{Krais2021}, together with likewise ordered side connectivity information and the grid coordinates. PICLas initially determines the number of elements per processor taking available load balancing information into account. Each processor only accesses mesh information for its region along the SFC and stores it in processor-local memory, thereby maximizing file system parallelism. However, each compute node additionally allocates shared memory sufficient to hold the raw mesh information (as stored in the HDF5 file) and save its compute node information at the correct offset. Once every processor has finished the reading of the mesh, the compute node root processors perform an non-blocking \verb!IALLGATHERV! operation on the interconnect, making use of available hardware offload capabilities. A graphical representation of this procedure is shown in~\cref{fig:impl:halo:readin}.

In addition to the unstructured computation grid, a Cartesian background mesh is created during runtime in order to reduce the eligible computational elements when performing particle localization procedures. The number of Cartesian elements in each direction is case-dependent and currently determined by the user. During runtime, particle intersection calculations are performed using either the physical element face corner coordinates in the case of a purely linear mesh or clipped Bézier surfaces for a curved grid. Additional particle information includes e.g., the face normal vectors, the distance of an element to the nearest boundary, the surrounding mesh node indices, and the tolerance of a curved element in reference space. Some of this information is only calculated given a specific tracking method and whether deposition is desired. Since part of this information varies depending on the mesh distribution and storing it would also inflate the size of the mesh file, these particle metrics are computed during the initialization phase.

\subsection{Halo Element Search Algorithm}
\label{sec:implementation:halo}
\begin{figure}[htb!]
\begin{subfigure}[t]{.5\textwidth}
\vspace{0pt}
\begin{adjustbox}{width=\textwidth}
\def\NodeSize{6pt}
\begin{tikzpicture}[]
  \pgfmathsetseed{6}%

  \coordinate (a1) at (0.0 , 0.0);
  \coordinate (a2) at (1.0 , 0.1);
  \coordinate (a3) at (1.1 , 1.2);
  \coordinate (a4) at (0.1 , 0.9);
  \draw[black,opacity=0.5,thick] (a1) -- (a2) -- (a3) -- (a4) -- cycle;
  \draw[fill=black,opacity=0.8] (a1) -- (a2) -- (a3) -- (a4) -- cycle;

  \coordinate (b1) at (0.9 , 2.0);
  \coordinate (b2) at (-.1 , 1.7);
  \draw[black,opacity=0.5,thick] (a3) -- (b1) -- (b2) -- (a4) -- cycle;
  \draw[fill=black,opacity=0.8] (a3) -- (b1) -- (b2) -- (a4) -- cycle;

  \coordinate (c1) at (1.9 , 1.9);
  \coordinate (c2) at (2.1 , 1.1);
  \draw[black,opacity=0.5,thick] (a3) -- (c2) -- (c1) -- (b1) -- cycle;
  \draw[fill=black,opacity=0.8] (a3) -- (c2) -- (c1) -- (b1) -- cycle;

  \coordinate (d1) at (2.0 , 0.0);
  \draw[black,opacity=0.5,thick] (a3) -- (c2) -- (d1) -- (a2) -- cycle;

  \coordinate (e1) at (2.9 , -.1);
  \coordinate (e2) at (3.0 , 1.1);
  \draw[black,opacity=0.5,thick] (d1) -- (e1) -- (e2) -- (c2) -- cycle;

  \coordinate (f1) at (3.0 , 2.1);
  \draw[black,opacity=0.5,thick] (e2) -- (f1) -- (c1) -- (c2) -- cycle;

  \coordinate (g1) at (2.9 , 3.0);
  \coordinate (g2) at (2.0 , 2.9);
  \draw[black,opacity=0.5,thick] (f1) -- (g1) -- (g2) -- (c1) -- cycle;

  \coordinate (h1) at (1.1 , 3.0);
  \draw[black,opacity=0.5,thick] (g2) -- (h1) -- (b1) -- (c1) -- cycle;
  \draw[fill=black,opacity=0.8] (g2) -- (h1) -- (b1) -- (c1) -- cycle;

  \coordinate (i1) at (0.1 , 2.9);
  \draw[black,opacity=0.5,thick] (h1) -- (i1) -- (b2) -- (b1) -- cycle;

  \coordinate (j1) at (-1.0 , 3.0);
  \coordinate (j2) at (-1.1 , 2.1);
  \draw[black,opacity=0.5,thick] (i1) -- (j1) -- (j2) -- (b2) -- cycle;

  \coordinate (k1) at (-1.0 , 1.0);
  \draw[black,opacity=0.5,thick] (j2) -- (k1) -- (a4) -- (b2) -- cycle;

  \coordinate (l1) at (-1.0 , 0.0);
  \draw[black,opacity=0.5,thick] (l1) -- (a1) -- (a4) -- (k1) -- cycle;

  \coordinate(y-1) at (-2.0+rand/5, -1.0+rand/5);
  \coordinate(y-2) at (-1.0+rand/5, -1.0+rand/5);
  \coordinate(y-3) at ( 0.0+rand/5, -1.0+rand/5);
  \coordinate(y-4) at ( 1.0+rand/5, -1.0+rand/5);
  \coordinate(y-5) at ( 2.0+rand/5, -1.0+rand/5);
  \coordinate(y-6) at ( 3.0+rand/5, -1.0+rand/5);

  \coordinate(x+1) at ( 4.0+rand/5, -1.0+rand/5);
  \coordinate(x+2) at ( 4.0+rand/5,  0.0+rand/5);
  \coordinate(x+3) at ( 4.0+rand/5,  1.0+rand/5);
  \coordinate(x+4) at ( 4.0+rand/5,  2.0+rand/5);
  \coordinate(x+5) at ( 4.0+rand/5,  3.0+rand/5);
  \coordinate(x+6) at ( 4.0+rand/5,  4.0+rand/5);

  \coordinate(y+1) at ( 3.0+rand/5,  4.0+rand/5);
  \coordinate(y+2) at ( 2.0+rand/5,  4.0+rand/5);
  \coordinate(y+3) at ( 1.0+rand/5,  4.0+rand/5);
  \coordinate(y+4) at ( 0.0+rand/5,  4.0+rand/5);
  \coordinate(y+5) at (-1.0+rand/5,  4.0+rand/5);
  \coordinate(y+6) at (-2.0+rand/5,  4.0+rand/5);

  \coordinate(x-1) at (-2.0+rand/5,  3.0+rand/5);
  \coordinate(x-2) at (-2.0+rand/5,  2.0+rand/5);
  \coordinate(x-3) at (-2.0+rand/5,  1.0+rand/5);
  \coordinate(x-4) at (-2.0+rand/5,  0.0+rand/5);

  \draw[black,opacity=0.5,thick] (y-1) -- (y-2) -- (l1) -- (x-4) -- cycle;
  \draw[black,opacity=0.5,thick] (y-2) -- (y-3) -- (a1) -- (l1)  -- cycle;
  \draw[black,opacity=0.5,thick] (y-3) -- (y-4) -- (a2) -- (a1)  -- cycle;
  \draw[black,opacity=0.5,thick] (y-4) -- (y-5) -- (d1) -- (a2)  -- cycle;
  \draw[black,opacity=0.5,thick] (y-5) -- (y-6) -- (e1) -- (d1)  -- cycle;

  \draw[black,opacity=0.5,thick] (y-6) -- (x+1) -- (x+2)-- (e1) -- cycle;
  \draw[black,opacity=0.5,thick] (x+2) -- (x+3) -- (e2) -- (e1) -- cycle;
  \draw[black,opacity=0.5,thick] (x+3) -- (x+4) -- (f1) -- (e2) -- cycle;
  \draw[black,opacity=0.5,thick] (x+4) -- (x+5) -- (g1) -- (f1) -- cycle;
  \draw[black,opacity=0.5,thick] (x+5) -- (x+6) -- (y+1) -- (g1) -- cycle;

  \draw[black,opacity=0.5,thick] (y+1) -- (y+2) -- (g2)  -- (g1) -- cycle;
  \draw[black,opacity=0.5,thick] (y+2) -- (y+3) -- (h1)  -- (g2) -- cycle;
  \draw[black,opacity=0.5,thick] (y+3) -- (y+4) -- (i1)  -- (h1) -- cycle;
  \draw[black,opacity=0.5,thick] (y+4) -- (y+5) -- (j1)  -- (i1) -- cycle;
  \draw[black,opacity=0.5,thick] (y+5) -- (y+6) -- (x-1) -- (j1) -- cycle;

  \draw[black,opacity=0.5,thick] (x-1) -- (x-2) -- (j2) -- (j1) -- cycle;
  \draw[black,opacity=0.5,thick] (x-2) -- (x-3) -- (k1) -- (j2) -- cycle;
  \draw[black,opacity=0.5,thick] (x-3) -- (x-4) -- (l1) -- (k1) -- cycle;

  \draw[black,opacity=0.5,thick] (-3,-2) -- (-2,-2) -- (y-1) -- (-3,-1) -- cycle;
  \draw[black,opacity=0.5,thick] (-2,-2) -- (-1,-2) -- (y-2) -- (y-1) -- cycle;
  \draw[black,opacity=0.5,thick] (-1,-2) -- ( 0,-2) -- (y-3) -- (y-2) -- cycle;
  \draw[black,opacity=0.5,thick] ( 0,-2) -- ( 1,-2) -- (y-4) -- (y-3) -- cycle;
  \draw[black,opacity=0.5,thick] ( 1,-2) -- ( 2,-2) -- (y-5) -- (y-4) -- cycle;
  \draw[black,opacity=0.5,thick] ( 2,-2) -- ( 3,-2) -- (y-6) -- (y-5) -- cycle;
  \draw[black,opacity=0.5,thick] ( 3,-2) -- ( 4,-2) -- (x+1) -- (y-6) -- cycle;
  \draw[black,opacity=0.5,thick] ( 4,-2) -- ( 5,-2) -- ( 5,-1) -- (x+1) -- cycle;

  \draw[black,opacity=0.5,thick] ( 5,-1) -- ( 5, 0) -- (x+2) -- (x+1) -- cycle;
  \draw[black,opacity=0.5,thick] ( 5, 0) -- ( 5, 1) -- (x+3) -- (x+2) -- cycle;
  \draw[black,opacity=0.5,thick] ( 5, 1) -- ( 5, 2) -- (x+4) -- (x+3) -- cycle;
  \draw[black,opacity=0.5,thick] ( 5, 2) -- ( 5, 3) -- (x+5) -- (x+4) -- cycle;
  \draw[black,opacity=0.5,thick] ( 5, 3) -- ( 5, 4) -- (x+6) -- (x+5) -- cycle;
  \draw[black,opacity=0.5,thick] ( 5, 4) -- ( 5, 5) -- ( 4, 5) -- (x+6) -- cycle;

  \draw[black,opacity=0.5,thick] ( 4, 5) -- ( 3, 5) -- (y+1) -- (x+6) -- cycle;
  \draw[black,opacity=0.5,thick] ( 3, 5) -- ( 2, 5) -- (y+2) -- (y+1) -- cycle;
  \draw[black,opacity=0.5,thick] ( 2, 5) -- ( 1, 5) -- (y+3) -- (y+2) -- cycle;
  \draw[black,opacity=0.5,thick] ( 1, 5) -- ( 0, 5) -- (y+4) -- (y+3) -- cycle;
  \draw[black,opacity=0.5,thick] ( 0, 5) -- (-1, 5) -- (y+5) -- (y+4) -- cycle;
  \draw[black,opacity=0.5,thick] (-1, 5) -- (-2, 5) -- (y+6) -- (y+5) -- cycle;
  \draw[black,opacity=0.5,thick] (-2, 5) -- (-3, 5) -- (-3, 4) -- (y+6) -- cycle;

  \draw[black,opacity=0.5,thick] (-3, 4) -- (-3, 3) -- (x-1) -- (y+6) -- cycle;
  \draw[black,opacity=0.5,thick] (-3, 3) -- (-3, 2) -- (x-2) -- (x-1) -- cycle;
  \draw[black,opacity=0.5,thick] (-3, 2) -- (-3, 1) -- (x-3) -- (x-2) -- cycle;
  \draw[black,opacity=0.5,thick] (-3, 1) -- (-3, 0) -- (x-4) -- (x-3) -- cycle;
  \draw[black,opacity=0.5,thick] (-3, 0) -- (-3,-1) -- (y-1) -- (x-4) -- cycle;



	\begin{pgfonlayer}{background}
    \foreach \x in {-3,-2.5,...,4.5} \foreach \y in {-2,-1.5,...,4.5}
    {
        \pgfmathparse{mod(\x+\y,2) ? "black!5" : "black!0"}
        \edef\colour{\pgfmathresult}
        \path[draw=black!20,fill=\colour] (\x,\y) rectangle ++ (0.5,0.5);
    };

    \begin{scope}[even odd rule]
      \draw[black, thick] (-1,-0.5) rectangle (3,3.5);
      \draw[fill=black,pattern=north west lines,opacity=0.8] (-1,-0.5) rectangle (3,3.5) (-0.1,0) rectangle (2.1,3);

      \begin{pgfinterruptboundingbox}
        \clip (-1,-0.5) rectangle (3,3.5) (-0.1,0) rectangle (2.1,3);
      \end{pgfinterruptboundingbox}
    \end{scope}
  \end{pgfonlayer}

  \draw[black, ultra thick] (-0.1,0) rectangle (2.1,3);
  \draw[fill=black,opacity=0.3] (-0.1,0) rectangle (2.1,3);

  \draw[black,thick] (2.5,-1.5) rectangle (4.5,0);
  \draw[fill=black,opacity=0.20] (2.5,-1.5) rectangle (4.5,0);
  \draw[fill=black,pattern=crosshatch dots,opacity=1.] (y-6) -- (x+1) -- (x+2)-- (e1) -- cycle;

  \coordinate(H1u) at ($(y-6)!0.5!(x+1)$);
  \coordinate(H1l) at ($(x+2)!0.5!(e1)$);
  \coordinate(H1)  at ($(H1u)!0.5!(H1l)$);
  \node[anchor=west] (H1s) at (3,-2.5) {Potential halo element};
  \draw[->,>=stealth,thick] (H1s) -- (H1);

  \draw[fill=black,pattern=north east lines,opacity=0.8] (2.5,-0.5) rectangle (3,0);

  \coordinate(O1)  at (2.75,-0.25);
  \node[anchor=west] (O1s) at (-1,-2.5) {BGM cell overlap};
  \draw[->,>=stealth,thick] (O1s) -- (O1);

  \begin{axis}[%
  hide axis,
  xmin=10,xmax=50,ymin=0,ymax=0.4,
  legend style={draw=white!15!black,legend cell align=left}
  ]
  \addlegendimage{fill=black,opacity=0.861,mark=none,area legend}
  \addlegendentry{Local mesh cells};
  \addlegendimage{black,pattern=north west lines,mark=none,area legend}
  \addlegendentry{Halo BGM bounding box};
  \end{axis}

\end{tikzpicture}
\end{adjustbox}
\caption{Identification of potential halo cells.}
\label{fig:halo:search:bgm}
\end{subfigure}
\begin{subfigure}[t]{.5\textwidth}
\vspace{0pt}
\begin{adjustbox}{width=\textwidth}
\def\NodeSize{6pt}
\input{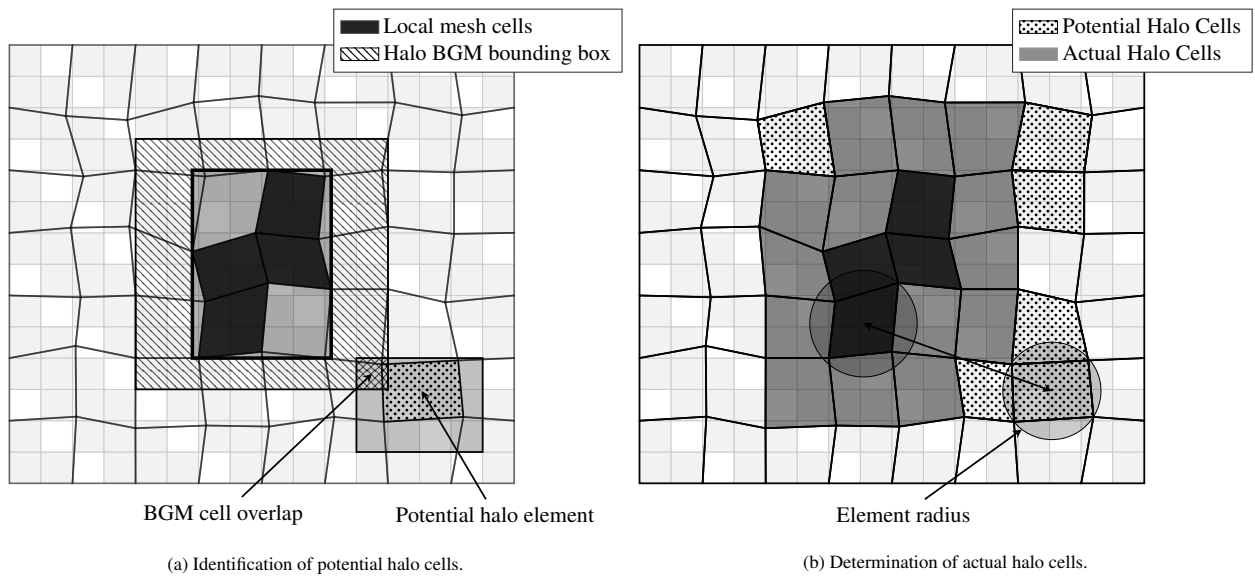}
\end{adjustbox}
\caption{Determination of actual halo cells.}
\label{fig:halo:search:distance}
\end{subfigure}
\caption{Determination of actual halo elements. Halo distance is \num{1} BGM cell.}
\end{figure}

In order to minimize both computational effort and memory requirements, a two-step search algorithm to determine eligible halo elements is performed before calculating the particle mesh metrics. This approach not only alleviates computational effort during the initialization phase but also significantly reduces the memory footprint as the derived metrics are only stored for the local and actual halo elements. For the first step depicted in \cref{fig:halo:search:bgm}, a Cartesian bounding box around all mesh elements local to a compute node is calculated. This bounding box is then extended by the halo distance in each direction.Through projection of the bounding box onto the Cartesian background mesh (BGM), the corresponding limits for the required I,J,K indices are obtained.

Next, a similar Cartesian bounding box is created for each mesh element. This bounding box is again projected onto the background mesh, thus creating a mapping from each mesh element to the overlapping BGM cells. The BGM mapping of every element not located on the compute node is then compared against the BGM region previously extended by the halo distance. All elements whose BGM bounding boxes overlap with the extended bounding box are flagged as potential halo elements. Since the global mesh information is available in the MPI-3 shared memory array, this calculation is distributed among all compute node processors through slicing of the space-filling curve.

At this point, the potential halo elements need to be further reduced as the Cartesian bounding box arbitrarily extends beyond the compute node local mesh elements. However, comparing the distance of every potential halo elements again all available compute node local elements would measurably affect the initialization time. Thus, the number of elements to compare against is reduced by only considering  the elements having at least one MPI boundary on the compute node circumference, therefore representing the MPI boundary of the compute node local mesh.

The second step shown in~\cref{fig:halo:search:distance} then calculates the radius of the convex hull of the elements on the MPI boundary and the radius of the potential halo elements. Each potential halo element is compared against each MPI-border element by subtracting the distance of the two barycenters from the sum of the two radii plus the halo distance. If the result is negative, the element is flagged as confirmed halo element. A positive value indicates that particles cannot reach the element within a given time increment and the element is subsequently discarded. This process is again distributed among all compute node processors through a uniform partitioning of the potential halo elements.

\begin{figure}[htb!]\centering
  \begin{subfigure}[c]{0.5\textwidth}\centering
    \begin{tikzpicture}

\definecolor{nodeelem}{RGB}{ 81, 81, 81}
\definecolor{nodegrid}{RGB}{218,218,218}
\definecolor{haloelem}{RGB}{185,185,185}
\definecolor{perielem}{RGB}{ 76, 76, 76}

\begin{axis}[
  xmin=0,xmax=100,
  ymin=0,ymax=100,
  hide axis,
  legend style={draw=white!15!black,legend cell align=left,at={(axis cs:105,105)},anchor=north east}
  ]
  \addlegendimage{area legend,preaction={fill=nodeelem},pattern=grid,pattern color=nodegrid};
  \addlegendentry{Local element};
  \addlegendimage{area legend,fill=haloelem}
  \addlegendentry{Halo element};
  \addlegendimage{area legend,fill=perielem}
  \addlegendentry{Periodic element};

  \addplot[] graphics[xmin=0,xmax=100,ymin=0,ymax=100] {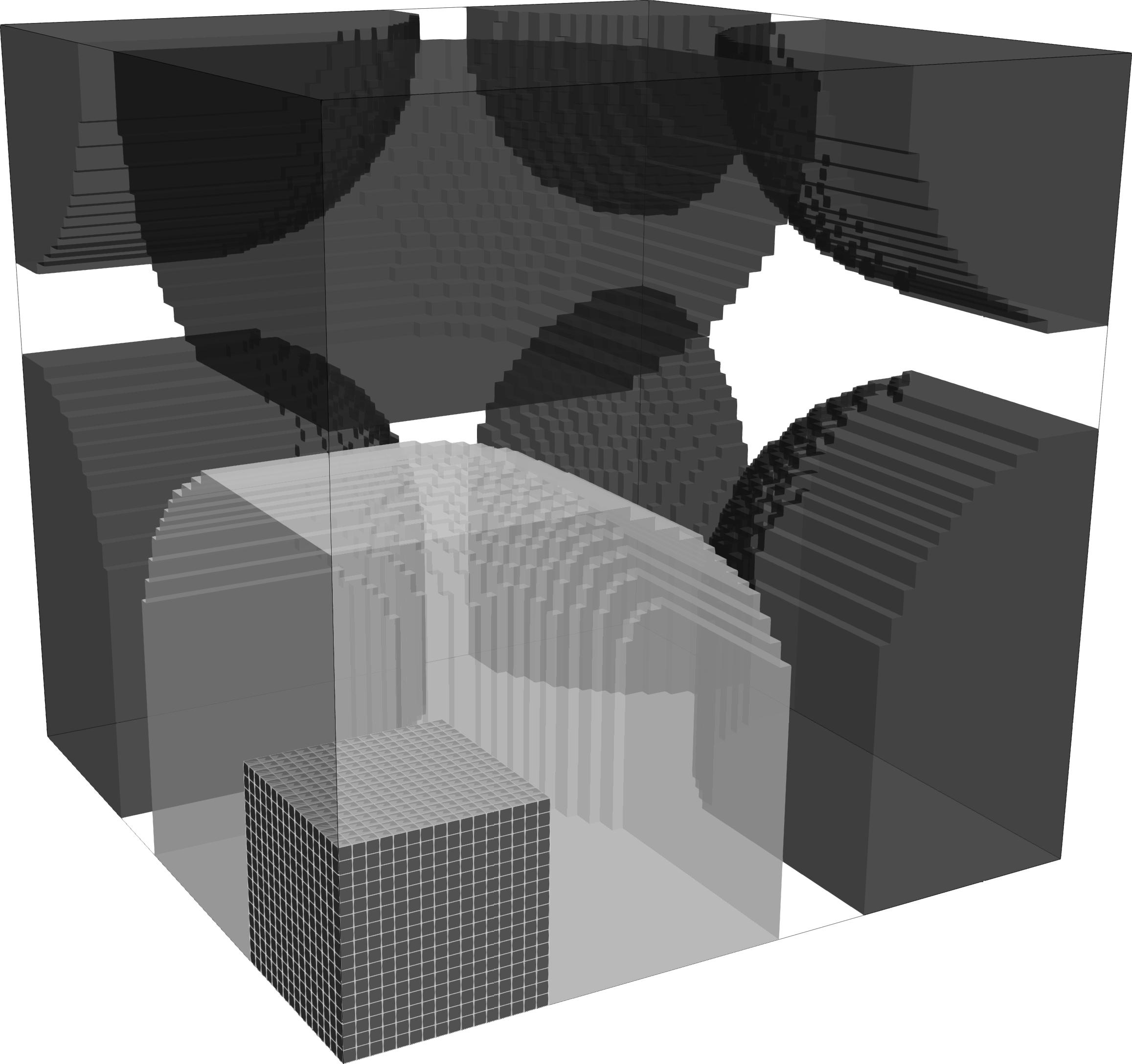};
\end{axis}
\end{tikzpicture}
  \end{subfigure}
  \caption{Local, halo and periodic element within halo distance for the adiabatic box, see~\cref{:sec:testcases:adiabaticbox}.}
  \label{fig:halo:search:flag}
\end{figure}

Only if the mesh contains periodic boundary conditions, the final step of the search is executed. In this case, all elements not yet flagged are moved with the periodic displacement vector and the distance evaluation depicted in~\cref{fig:halo:search:distance} is performed again. If the periodically displaced element is within the halo distance, it is added to the halo cells with an indicator to consider this element only for tracking purposes. An example of flagged elements for the fully periodic adiabatic box, see also~\cref{:sec:testcases:adiabaticbox}, is shown in~\cref{fig:halo:search:flag}.

Once every compute node processor indicates that there are no more potential halo elements to check, a mapping containing first the compute node local elements followed by the compute node halo elements is built. This mapping allows for efficient looping when building the derived particle mesh metrics on the reduced mesh.

\subsection{Emission}
\label{sec:implementation:emission}
Since particles are tracked in physical space, this approach naturally has to extend to the particle emission as well. In order to maintain good scalability, emission is performed in parallel with each processor calculating the initial particle positions within the complete emission region. Subsequently, the grid element corresponding to each position has to be identified and the particle sent to the respective processor. However, the emission region might extend beyond the compute node-local mesh even when including the halo region, meaning that the emitting processor cannot uniquely identify the target element. Worse yet, without knowledge of the elements outside the halo region, the emitting processor cannot distinguish between a valid position and one outside of the complete mesh. Yet, it is equally undesirable to retain all elements within the emission region.

The solution to the problem is again provided through the BGM. As mentioned in the previous paragraph, there exists a mapping from each BGM element to the overlapping grid elements located within the compute node and halo region. During initialization, each processor additionally provides the number of local elements per BGM cell. This number is summed up across all processors and stored with a mapping containing all \textit{processors} which overlap with a given BGM cell. As the local elements are inherently distributed, this process is by design automatically scaling.

During the particle emission, after an emission position is computed, the processor calculates the associated BGM cell. Next, we compare the number of compute node grid elements mapped to this BGM cell with the total number of grid elements connected to this cell. If the numbers match, the processor flags the position as locally computable. The other positions are gathered and sent to all processors associated with the BGM cell. Next, we perform the search algorithm of the locally computable positions, thereby acting as latency hiding. After identifying all local particle to element mappings, the search of communicated positions is performed on each processor. Since any position can only correspond to one single element, no further communication is required. All other processors silently discard the position.

\subsection{Deposition with Shape Function}
\label{sec:implementation:shape}
\begin{figure}[htb!]
\centering\begin{adjustbox}{width=.5\textwidth}
\def\NodeSize{6pt}
\input{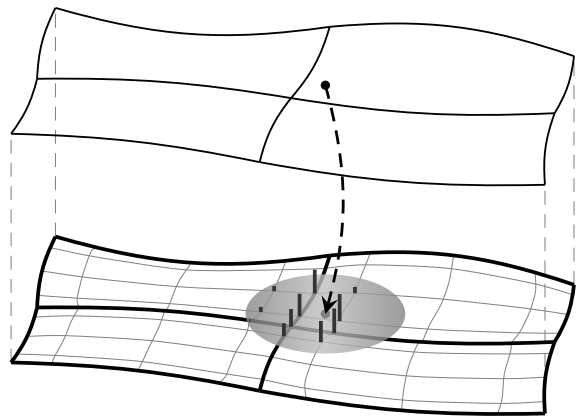}
\end{adjustbox}
\caption{Charge deposition on degree of freedom.}
\label{fig:impl:shape}
\end{figure}

The charged particles are responsible for the source terms of the field equations and therefore are coupled with the underlying grid on which the field equations are solved. The source terms themselves are determined from the respective moments as described by the distribution function. Here, this is achieved by mapping the particle position and velocity to the grid via shape functions that smoothly distribute the charge and current densities of the particles on the grid, which is referred to as deposition. The cut-off radius of the deposition is determined by the physical problem and can range across multiple elements of the grid as shown in~\cref{fig:impl:shape}, hence, deposition may occur in processor-local elements as well as elements that are of different processors or even different nodes. Thus, processors require the communication of either the deposited properties or the particle properties, which are in turn deposited by the receiving processor. 
Since the presented parallelization concept allows elements to be uniquely globally identified, the particles are deposited by the host process of the particle. Source terms that are possibly deposited in other processors are stored in a separate array for communication. Subsequently, a message is created from this array for all processors in whose elements were deposited and communicated regardless of which node they are located on. In order to communicate only with processors that can potentially exchange source terms, a list of all reachable processors in the halo region is initially created using the shape function radius for each processor. In order to avoid multiple communications between all processors in this initial process, this is done in a two-step communication procedure. In a first step, the list with the processors to be communicated is sent to the MPI roots of the compute node for each processor. These node leaders on each compute node gather the information for distribution to other processors from all corresponding processors and store them in a shared array so that the information about the necessary exchange processors is available to all processors.

\subsection{Latency Hiding}
The main goal of latency hiding is to allow communication and computation to overlap completely, which means that during the time of communication, parts of the algorithm are already being carried out, so that there is no waiting time during communication.
There are two basic problems with latency hiding for PICLas. PICLas is a modular toolbox, so e.g., PIC can be used as a module without or with collision term, within PIC there is the distinction whether electromagnetic or electrostatic simulations should be carried out or which interpolation method between particles and grid should be used. On the collision term side, there is also a wide choice of methods such as DSMC, BGK, FP and others. Each possible combination of modules with different types of time integration has different requirements as to which data must be available and when. Obviously, it is therefore not possible to use a latency hiding method that represents the optimum for all possible module combinations. Therefore, in the following we will only concentrate on the methods used in this work and already described in the theory section. The other problem is that many particle methods such as DSMC, BGK or FP have very sequential structures, so that only a few parts can already be calculated if the information from particles from other cores is not yet available. Most of the time, calculations that are carried out in parallel with communication require a large amount of additional memory, because quantities that would otherwise only be needed locally in the cell have to be stored for all cells so that they are available after communication.

\subsubsection{Maxwell-PIC with Shape Functions}
Maxwell-PIC within this context refers to the electrodynamic PIC method in which the complete set of Maxwell's equations are solved as described in \cref{sec:theory:Plasma}. 
The time integration scheme used in this example is a 5-stage $4^\text{th}$-order low-storage Runge-Kutta (RK) method~\cite{carpenter1994fourth}. For simulations of this type, there are three MPI communications per time step which should require latency masking, i.e., the communication of the flux data for the Maxwell-DG solver, the current densities and charge densities deposited on the grid as source terms for the DG solver by the shape functions and the particle data that leave the processor after the movement. In general, the compute time of the discrete phase is dominating compared to the continuous phase, hence the focus of latency hiding is to avoid stalling of the particle routines.

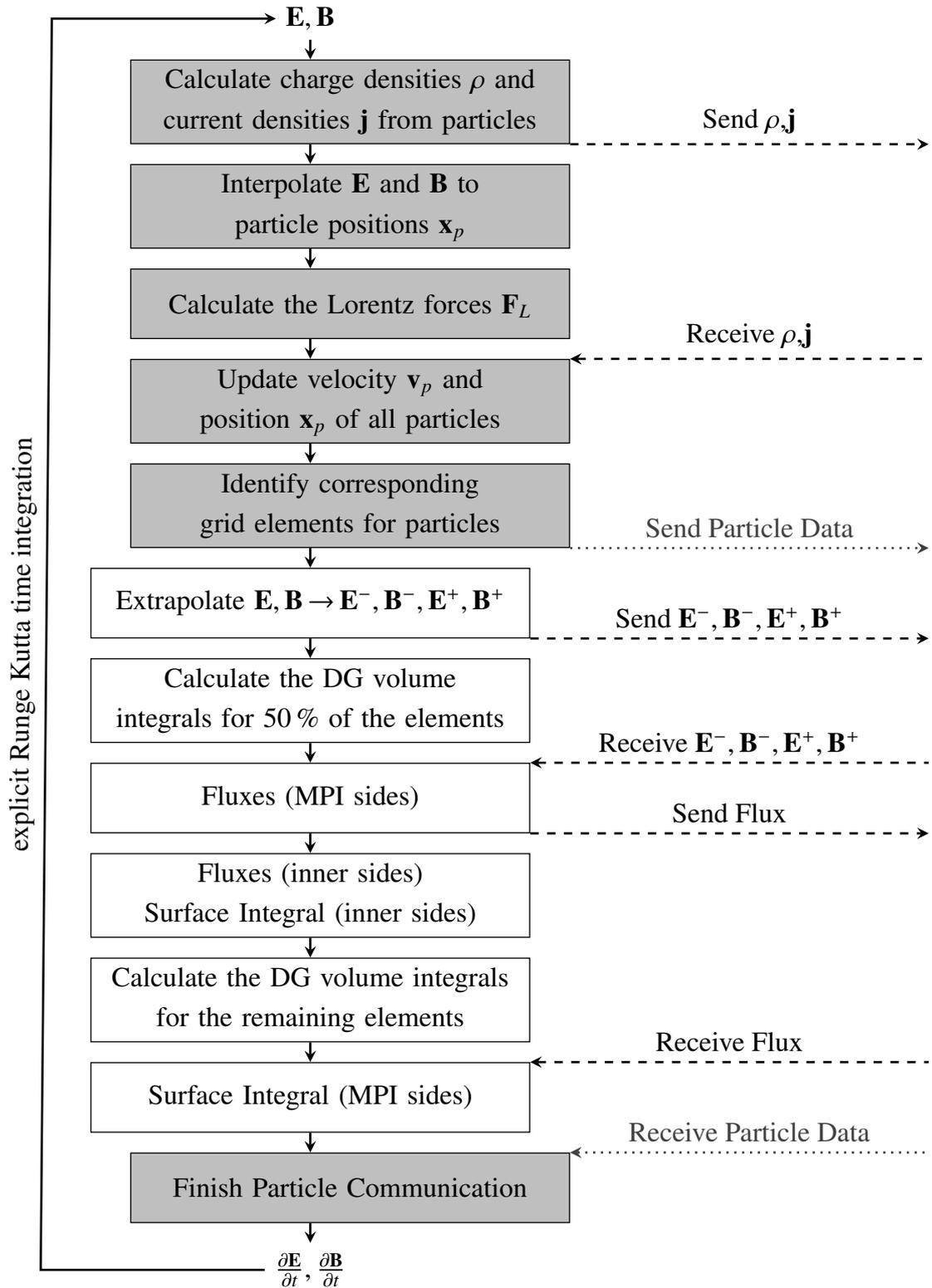
\begin{figure}[htbp!]
  \centering\begin{adjustbox}{width=0.95\textwidth}
\def\NodeSize{6pt}
\tikzstyle{steps} = [rectangle, minimum width=14em, minimum height=2.5em, text centered, text width=15em, draw=black, fill=white, anchor=north]

\setlength{\abovedisplayskip}{0pt}
\setlength{\belowdisplayskip}{0pt}
\setlength{\abovedisplayshortskip}{0pt}
\setlength{\belowdisplayshortskip}{0pt}
\renewcommand{\baselinestretch}{1.2}

\begin{tikzpicture}[node distance=1.5em]
  \node (start) []                                                       {$\mathbf{E},\mathbf{B}$};
  \node (part1) [steps,fill=gray!50] at ($(start.south) + ( 0.5,-0.25)$) {Calculate charge densities $\smash{\rho}$ and current densities $\smash{\mathbf{j}}$ from particles};
  \node (part2) [steps,fill=gray!50] at ($(part1.south) + ( 0.0,-0.25)$) {Interpolate $\mathbf{E}$ and $\mathbf{B}$ to particle positions $\smash{\mathbf{x}_p}$};
  \node (part3) [steps,fill=gray!50] at ($(part2.south) + ( 0.0,-0.25)$) {Calculate the Lorentz forces $\smash{\mathbf{F}_L}$};
  \node (part4) [steps,fill=gray!50] at ($(part3.south) + ( 0.0,-0.25)$) {Update velocity $\smash{\mathbf{v}_p}$ and position $\smash{\mathbf{x}_p}$ of all particles};
  \node (part5) [steps,fill=gray!50] at ($(part4.south) + ( 0.0,-0.25)$) {Identify corresponding grid elements for particles};
  \node (step1) [steps             ] at ($(part5.south) + (-0.5,-0.25)$) {Extrapolate $\smash{\mathbf{E},\mathbf{B}\,{\rightarrow}\,\mathbf{E}^{-},\mathbf{B}^{-},\mathbf{E}^{+},\mathbf{B}^{+}}$};
  \node (step2) [steps             ] at ($(step1.south) + ( 0.0,-0.25)$) {Calculate the DG volume integrals for \SI{50}{\%} of the elements};
  \node (step3) [steps,align=center] at ($(step2.south) + ( 0.0,-0.25)$) {Fluxes (MPI sides)};
  \node (step4) [steps,align=center] at ($(step3.south) + ( 0.0,-0.25)$) {Fluxes (inner sides)\\ Surface Integral (inner sides)};
  \node (step5) [steps             ] at ($(step4.south) + ( 0.0,-0.25)$) {Calculate the DG volume integrals for the remaining elements};
  \node (step6) [steps,align=center] at ($(step5.south) + ( 0.0,-0.25)$) {Surface Integral (MPI sides)};
  \node (part6) [steps,fill=gray!50] at ($(step6.south) + ( 0.5,-0.25)$) {Finish Particle Communication};
  \node (end)   [anchor=north]       at ($(part6.south) + (-0.5,-0.25)$) {$\frac{\partial\mathbf{E}}{\partial t},\frac{\partial \mathbf{B}}{\partial t}$};

  \draw[->,thick,>=stealth]   (start.south)              -- ($(part1.north)+(-0.5,0.0)$);
  \draw[->,thick,>=stealth] ($(part1.south)+(-0.5,0.0)$) -- ($(part2.north)+(-0.5,0.0)$);
  \draw[->,thick,>=stealth] ($(part2.south)+(-0.5,0.0)$) -- ($(part3.north)+(-0.5,0.0)$);
  \draw[->,thick,>=stealth] ($(part3.south)+(-0.5,0.0)$) -- ($(part4.north)+(-0.5,0.0)$);
  \draw[->,thick,>=stealth] ($(part4.south)+(-0.5,0.0)$) -- ($(part5.north)+(-0.5,0.0)$);
  \draw[->,thick,>=stealth] ($(part5.south)+(-0.5,0.0)$) --   (step1.north);
  \draw[->,thick,>=stealth]   (step1.south)              --   (step2.north);
  \draw[->,thick,>=stealth]   (step2.south)              --   (step3.north);
  \draw[->,thick,>=stealth]   (step3.south)              --   (step4.north);
  \draw[->,thick,>=stealth]   (step4.south)              --   (step5.north);
  \draw[->,thick,>=stealth]   (step5.south)              --   (step6.north);
  \draw[->,thick,>=stealth]   (step6.south)              -- ($(part6.north)+(-0.5,0.0)$);
  \draw[->,thick,>=stealth] ($(part6.south)+(-0.5,0.0)$) --   (end.north);
  \draw[->,thick,>=stealth]   (end.west) -- ++(-8em,0)   -- node[midway,sloped,above] {explicit Runge Kutta time integration} ($(start.west)+(-8em,0)$) -- (start.west);

  \draw[->,>=stealth,thick,dashed     ]     (step1.south east)                          -- node[above] {Send    $\smash{\mathbf{E}^{-},\mathbf{B}^{-},\mathbf{E}^{+},\mathbf{B}^{+}}$}   ++(5.0,0) ;
  \draw[<-,>=stealth,thick,dashed     ]     (step3.north east)                          -- node[above] {Receive $\smash{\mathbf{E}^{-},\mathbf{B}^{-},\mathbf{E}^{+},\mathbf{B}^{+}}$}   ++(5.0,0) ;
  \draw[->,>=stealth,thick,dashed     ]     (step3.south east)                          -- node[above] {Send    Flux}            ++(5.0,0) ;
  \draw[<-,>=stealth,thick,dashed     ]     (step6.north east)                          -- node[above] {Receive Flux}            ++(5.0,0) ;

  \draw[->,>=stealth,thick,dashed     ]     (part1.south east)                          -- node[above] {Send    $\rho$,$\mathbf{j}$} ++(4.5,0) ;
  \draw[<-,>=stealth,thick,dashed     ]     (part4.north east)                          -- node[above] {Receive $\rho$,$\mathbf{j}$} ++(4.5,0) ;
  \draw[->,>=stealth,thick,dotted,black!75] (part5.south east)                          -- node[above] {Send    Particle Data}   ++(4.5,0) ;
  \draw[<-,>=stealth,thick,dotted,black!75] (part6.north east)                          -- node[above] {Receive Particle Data}   ++(4.5,0) ;
\end{tikzpicture}
\end{adjustbox}
\caption{Latency hiding for Maxwell-PIC with shape functions.}
\label{fig:impl:latency:pic}
\end{figure}

A flow chart of one RK stage is given in~\cref{fig:impl:latency:pic}. Performing the deposition as outlined in~\cref{sec:implementation:shape} allows to hide the costly exchange of volume data behind the particle operators. Interpolation, calculation of the Lorenz forces as well as particle tracking are purely local operations. At their end, particles are already assigned to their final processor, so the communication of particle data can start. As the continuous phase requires two communication steps, we start by extrapolating the field data to the element faces and communicate the surfaces data. This corresponds to $(N-)\text{D}$ information and thus requires considerably less interconnect time than the previous field data. As was already outlined, the DG volume integral is a purely local operation and is thus performed on the first half of the local elements to hide the communication latency. After receiving the surface data, the numerical flux on the MPI sides is calculated and immediately sent back. Local operations on the inner sides as well as the remaining volume integrals can commence. Once these routines return, the numerical flux on the MPI sides should be received and the surface integral on the MPI sides is calculated. The RK stage concludes by incorporating the particle data into the local arrays.

\subsubsection{SP-BGK}
\label{sec:LH_BGK}
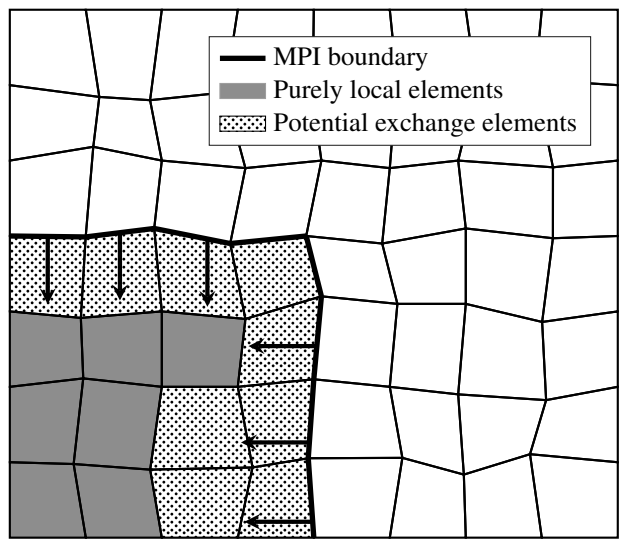
\begin{figure}[htb!]
\centering\begin{adjustbox}{width=.5\textwidth}
\def\NodeSize{6pt}

\begin{tikzpicture}[]
  \pgfmathsetseed{4}%

  \coordinate (a1) at (0.0 , 0.0);
  \coordinate (a2) at (1.0 , 0.1);
  \coordinate (a3) at (1.1 , 1.2);
  \coordinate (a4) at (0.1 , 0.9);
  \draw[black,thick] (a1) -- (a2) -- (a3) -- (a4) -- cycle;

  \coordinate (b1) at (0.9 , 2.0);
  \coordinate (b2) at (-.1 , 1.9);
  \draw[black,thick] (a3) -- (b1) -- (b2) -- (a4) -- cycle;

  \coordinate (c1) at (1.9 , 1.9);
  \coordinate (c2) at (2.1 , 1.1);
  \draw[black,thick] (a3) -- (c2) -- (c1) -- (b1) -- cycle;

  \coordinate (d1) at (2.0 , 0.0);
  \draw[black,thick] (a3) -- (c2) -- (d1) -- (a2) -- cycle;

  \coordinate (e1) at (2.9 , -.1);
  \coordinate (e2) at (3.0 , 1.1);
  \draw[black,thick] (d1) -- (e1) -- (e2) -- (c2) -- cycle;

  \coordinate (f1) at (3.0 , 2.1);
  \draw[black,thick] (e2) -- (f1) -- (c1) -- (c2) -- cycle;

  \coordinate (g1) at (2.9 , 3.0);
  \coordinate (g2) at (2.0 , 2.9);
  \draw[black,thick] (f1) -- (g1) -- (g2) -- (c1) -- cycle;

  \coordinate (h1) at (1.1 , 3.0);
  \draw[black,thick] (g2) -- (h1) -- (b1) -- (c1) -- cycle;

  \coordinate (i1) at (0.1 , 2.9);
  \draw[black,thick] (h1) -- (i1) -- (b2) -- (b1) -- cycle;

  \coordinate (j1) at (-1.0 , 3.0);
  \coordinate (j2) at (-1.1 , 2.1);
  \draw[black,thick] (i1) -- (j1) -- (j2) -- (b2) -- cycle;

  \coordinate (k1) at (-1.0 , 1.0);
  \draw[black,thick] (j2) -- (k1) -- (a4) -- (b2) -- cycle;

  \coordinate (l1) at (-1.0 , 0.0);
  \draw[black,thick] (l1) -- (a1) -- (a4) -- (k1) -- cycle;

  \coordinate(y-1) at (-2.0+rand/5, -1.0+rand/5);
  \coordinate(y-2) at (-1.0+rand/5, -1.0+rand/5);
  \coordinate(y-3) at ( 0.0+rand/5, -1.0+rand/5);
  \coordinate(y-4) at ( 1.0+rand/5, -1.0+rand/5);
  \coordinate(y-5) at ( 2.0+rand/5, -1.0+rand/5);
  \coordinate(y-6) at ( 3.0+rand/5, -1.0+rand/5);

  \coordinate(x+1) at ( 4.0+rand/5, -1.0+rand/5);
  \coordinate(x+2) at ( 4.0+rand/5,  0.0+rand/5);
  \coordinate(x+3) at ( 4.0+rand/5,  1.0+rand/5);
  \coordinate(x+4) at ( 4.0+rand/5,  2.0+rand/5);
  \coordinate(x+5) at ( 4.0+rand/5,  3.0+rand/5);
  \coordinate(x+6) at ( 4.0+rand/5,  4.0+rand/5);

  \coordinate(y+1) at ( 3.0+rand/5,  4.0+rand/5);
  \coordinate(y+2) at ( 2.0+rand/5,  4.0+rand/5);
  \coordinate(y+3) at ( 1.0+rand/5,  4.0+rand/5);
  \coordinate(y+4) at ( 0.0+rand/5,  4.0+rand/5);
  \coordinate(y+5) at (-1.0+rand/5,  4.0+rand/5);
  \coordinate(y+6) at (-2.0+rand/5,  4.0+rand/5);

  \coordinate(x-1) at (-2.0+rand/5,  3.0+rand/5);
  \coordinate(x-2) at (-2.0+rand/5,  2.0+rand/5);
  \coordinate(x-3) at (-2.0+rand/5,  1.0+rand/5);
  \coordinate(x-4) at (-2.0+rand/5,  0.0+rand/5);

  \draw[black,thick] (y-1) -- (y-2) -- (l1) -- (x-4) -- cycle;
  \draw[black,thick] (y-2) -- (y-3) -- (a1) -- (l1)  -- cycle;
  \draw[black,thick] (y-3) -- (y-4) -- (a2) -- (a1)  -- cycle;
  \draw[black,thick] (y-4) -- (y-5) -- (d1) -- (a2)  -- cycle;
  \draw[black,thick] (y-5) -- (y-6) -- (e1) -- (d1)  -- cycle;

  \draw[black,thick] (y-6) -- (x+1) -- (x+2)-- (e1) -- cycle;
  \draw[black,thick] (x+2) -- (x+3) -- (e2) -- (e1) -- cycle;
  \draw[black,thick] (x+3) -- (x+4) -- (f1) -- (e2) -- cycle;
  \draw[black,thick] (x+4) -- (x+5) -- (g1) -- (f1) -- cycle;
  \draw[black,thick] (x+5) -- (x+6) -- (y+1) -- (g1) -- cycle;

  \draw[black,thick] (y+1) -- (y+2) -- (g2)  -- (g1) -- cycle;
  \draw[black,thick] (y+2) -- (y+3) -- (h1)  -- (g2) -- cycle;
  \draw[black,thick] (y+3) -- (y+4) -- (i1)  -- (h1) -- cycle;
  \draw[black,thick] (y+4) -- (y+5) -- (j1)  -- (i1) -- cycle;
  \draw[black,thick] (y+5) -- (y+6) -- (x-1) -- (j1) -- cycle;

  \draw[black,thick] (x-1) -- (x-2) -- (j2) -- (j1) -- cycle;
  \draw[black,thick] (x-2) -- (x-3) -- (k1) -- (j2) -- cycle;
  \draw[black,thick] (x-3) -- (x-4) -- (l1) -- (k1) -- cycle;

  \draw[black,thick] (-3,-2) -- (-2,-2) -- (y-1) -- (-3,-1) -- cycle;
  \draw[black,thick] (-2,-2) -- (-1,-2) -- (y-2) -- (y-1) -- cycle;
  \draw[black,thick] (-1,-2) -- ( 0,-2) -- (y-3) -- (y-2) -- cycle;
  \draw[black,thick] ( 0,-2) -- ( 1,-2) -- (y-4) -- (y-3) -- cycle;
  \draw[black,thick] ( 1,-2) -- ( 2,-2) -- (y-5) -- (y-4) -- cycle;
  \draw[black,thick] ( 2,-2) -- ( 3,-2) -- (y-6) -- (y-5) -- cycle;
  \draw[black,thick] ( 3,-2) -- ( 4,-2) -- (x+1) -- (y-6) -- cycle;
  \draw[black,thick] ( 4,-2) -- ( 5,-2) -- ( 5,-1) -- (x+1) -- cycle;

  \draw[black,thick] ( 5,-1) -- ( 5, 0) -- (x+2) -- (x+1) -- cycle;
  \draw[black,thick] ( 5, 0) -- ( 5, 1) -- (x+3) -- (x+2) -- cycle;
  \draw[black,thick] ( 5, 1) -- ( 5, 2) -- (x+4) -- (x+3) -- cycle;
  \draw[black,thick] ( 5, 2) -- ( 5, 3) -- (x+5) -- (x+4) -- cycle;
  \draw[black,thick] ( 5, 3) -- ( 5, 4) -- (x+6) -- (x+5) -- cycle;
  \draw[black,thick] ( 5, 4) -- ( 5, 5) -- ( 4, 5) -- (x+6) -- cycle;

  \draw[black,thick] ( 4, 5) -- ( 3, 5) -- (y+1) -- (x+6) -- cycle;
  \draw[black,thick] ( 3, 5) -- ( 2, 5) -- (y+2) -- (y+1) -- cycle;
  \draw[black,thick] ( 2, 5) -- ( 1, 5) -- (y+3) -- (y+2) -- cycle;
  \draw[black,thick] ( 1, 5) -- ( 0, 5) -- (y+4) -- (y+3) -- cycle;
  \draw[black,thick] ( 0, 5) -- (-1, 5) -- (y+5) -- (y+4) -- cycle;
  \draw[black,thick] (-1, 5) -- (-2, 5) -- (y+6) -- (y+5) -- cycle;
  \draw[black,thick] (-2, 5) -- (-3, 5) -- (-3, 4) -- (y+6) -- cycle;

  \draw[black,thick] (-3, 4) -- (-3, 3) -- (x-1) -- (y+6) -- cycle;
  \draw[black,thick] (-3, 3) -- (-3, 2) -- (x-2) -- (x-1) -- cycle;
  \draw[black,thick] (-3, 2) -- (-3, 1) -- (x-3) -- (x-2) -- cycle;
  \draw[black,thick] (-3, 1) -- (-3, 0) -- (x-4) -- (x-3) -- cycle;
  \draw[black,thick] (-3, 0) -- (-3,-1) -- (y-1) -- (x-4) -- cycle;

  \draw[fill=black,opacity=0.45] (-3,-2) -- (-2,-2) -- (y-1) -- (-3,-1) -- cycle;
  \draw[fill=black,opacity=0.45] (-3, 0) -- (-3,-1) -- (y-1) -- (x-4) -- cycle;
  \draw[fill=black,opacity=0.45] (-3, 1) -- (-3, 0) -- (x-4) -- (x-3) -- cycle;
  \draw[black,pattern=crosshatch dots] (-3, 2) -- (-3, 1) -- (x-3) -- (x-2) -- cycle;

  \draw[fill=black,opacity=0.45] (-2,-2) -- (-1,-2) -- (y-2) -- (y-1) -- cycle;
  \draw[fill=black,opacity=0.45] (y-1) -- (y-2) -- (l1) -- (x-4) -- cycle;
  \draw[fill=black,opacity=0.45] (x-3) -- (x-4) -- (l1) -- (k1) -- cycle;
  \draw[black,pattern=crosshatch dots] (x-2) -- (x-3) -- (k1) -- (j2) -- cycle;

  \draw[black,pattern=crosshatch dots] (y-2) -- (y-3) -- (a1) -- (l1)  -- cycle;
  \draw[black,pattern=crosshatch dots] (-1,-2) -- ( 0,-2) -- (y-3) -- (y-2) -- cycle;
  \draw[fill=black,opacity=0.45] (l1) -- (a1) -- (a4) -- (k1) -- cycle;
  \draw[black,pattern=crosshatch dots] (j2) -- (k1) -- (a4) -- (b2) -- cycle;

  \draw[black,pattern=crosshatch dots] (y-3) -- (y-4) -- (a2) -- (a1)  -- cycle;
  \draw[black,pattern=crosshatch dots] ( 0,-2) -- ( 1,-2) -- (y-4) -- (y-3) -- cycle;
  \draw[black,pattern=crosshatch dots] (a1) -- (a2) -- (a3) -- (a4) -- cycle;
  \draw[black,pattern=crosshatch dots] (a3) -- (b1) -- (b2) -- (a4) -- cycle;

  \draw[->,>=stealth,ultra thick,black] ($(-3,2)!0.5!(x-2)$) -- ++( 0.00,-0.90);
  \draw[->,>=stealth,ultra thick,black] ($(j2)!0.5!(x-2)$)   -- ++( 0.00,-0.90);
  \draw[->,>=stealth,ultra thick,black] ($(j2)!0.7!(b2)$)    -- ++( 0.00,-0.90);
  \draw[->,>=stealth,ultra thick,black] ($(a2)!0.4!(a3)$)    -- ++(-0.90, 0.00);
  \draw[->,>=stealth,ultra thick,black] ($(y-4)!0.2!(a2)$)   -- ++(-0.90, 0.00);
  \draw[->,>=stealth,ultra thick,black] ($(1,-2)!0.2!(y-4)$) -- ++(-0.90, 0.00);

  \draw[black,line width=2pt] (-3, 2) -- (x-2);
  \draw[black,line width=2pt] (x-2)   -- (j2);
  \draw[black,line width=2pt] (j2)    -- (b2);
  \draw[black,line width=2pt] (b2)    -- (b1);
  \draw[black,line width=2pt] (b1)    -- (a3);
  \draw[black,line width=2pt] (a3)    -- (a2);
  \draw[black,line width=2pt] (a2)    -- (y-4);
  \draw[black,line width=2pt] (y-4)   -- ( 1,-2);

  \begin{axis}[%
  hide axis,
  height=180,width=180,
  xmin=0,xmax=1,ymin=0,ymax=1,
  legend style={draw=white!15!black,legend cell align=left}
  ]
  \addlegendimage{black,line width=2pt,mark=none}
  \addlegendentry{MPI boundary};
  \addlegendimage{fill=black,opacity=0.45,mark=none,area legend}
  \addlegendentry{Purely local elements};
  \addlegendimage{black,pattern=crosshatch dots,mark=none,area legend}
  \addlegendentry{Potential exchange elements};
  \end{axis}

\end{tikzpicture}
\end{adjustbox}
\caption{Latency hiding for the BGK method.}
\label{fig:impl:latency:bgk}
\end{figure}

The SP-BGK method requires the allocation of the particles to the cells in order to calculate the moments of the distribution function per cell. However, this is a purely discrete phase method, thus no field solver part or deposition to hide the communication time of the particle communication as in the PIC method.
Therefore, latency hiding is implemented here in two different approaches. In the first approach, the elements of each processor are divided into exchange elements in which particles from other processors can potentially move within a time step and purely local elements in which this is not possible. For this purpose, the halo region is extended from the MPI boundaries into each processor's own computational domain and the elements are flagged accordingly as shown in~\cref{fig:impl:latency:bgk}. With this information, the BGK operator can already be applied to the purely local elements during particle communication, as this operator is cell local. After all particles have been received, the BGK operator is then applied to the exchange elements. This type of latency hiding is only effective if each processor has enough elements so that a minimum amount of purely local elements exists. However, this is often not the case for very high numbers of processors. Therefore, the assignment of the particles to the elements is additionally divided. All particles that do not leave the processor's slice of the computational domain are already assigned to the elements during the particle communication. Within the elements, an adaptive octree is used to create subcells in order to better capture any gradients~\cite{pfeiffer2013grid,Fasoulas2019}. As far as possible, parts of this assignment are also carried out during the communication of the particles. Subsequently, after the particle data is received, the element assignment is done for the received particles. This second step of latency hiding has the advantage that it can always be performed regardless of the number of purely local elements.

\section{Test Cases}
\label{sec:testcases}
Three test cases were selected to investigate the efficiency of the proposed methods. The first test case consists of a generic setup representing a weak scaling case of an adiabatic periodic Cartesian box to assess specific performance metrics in a regular and adaptable setup. The second and third simulation showcase the strong scaling of practical application setups with fixed sizes on unstructured grids. This section presents the test cases themselves with the scaling results given in~\cref{sec:results}.

All simulations are performed on the HPE Apollo \hawk\ system at the High Performance Computing Center Stuttgart (HLRS) with dual-socket AMD EPYC\textsuperscript{TM} nodes featuring 128 cores per node and an InfiniBand HDR200 interconnect. The interconnect is deployed in a \num{9}-dimensional enhanced hypercube topology. Within this topology, \num{16} nodes are directly connected to a switch, providing the full \SI{200}{Gbit/s} bandwidth. A dotted vertical line is shown in each plot to indicate the initial bandwidth drop. Higher dimensions, i.e., increasing node numbers, result in decreasing interconnect bandwidth and growing latency~\cite{hlrscube2020}. The code was compiled with the GNU compiler version 9.2.0 with the libraries mpt 2.23, hdf5 1.10.5 and aocl 3.0. Each run was repeated three or more times to eliminate fluctuations in overall machine load and bandwidth contentions.

\subsection{Adiabatic Box}
\label{:sec:testcases:adiabaticbox}
The first test is an adiabatic box which represents the optimum for the parallelization. The domain itself is a fully-periodic 3D box into which methane with a particle density of $n=\SI{1E23}{1/m^3}$ is homogeneously inserted. The start temperature is $T_\infty=\SI{2000}{K}$, resulting in a required time step of $\Delta t =\SI{3E-10}{s}$~\cite{pfeiffer2018particle}. The simulation is carried out using the stochastic particle Bhatnagar-Gross-Krook (SP-BGK) method as described in \cref{sec:theory:BGK}, with the periodic structure of the domain ensuring that the number of particles remains homogeneous in the entire computational domain and thus the computational load for all processors is equally distributed over the entire computation time. Nevertheless, of course, the particles move at each time step and the collision operator is also executed.

As the physical dimensions of the cubic domain can easily be adjusted, this test case is utilized to investigate the weak scaling behavior for an ideal case. The elementary unit for the weak scaling represents a computational domain of $16\times 16 \times 16$ cells with an edge length of $\SI{50E-5}{m}$ each. This results in \num{32} elements per core on a node with \num{128} cores. If the number of nodes is doubled, the domain size is also doubled to achieve the same number of elements per core, i.e. \num{2} nodes results in $32\times 16 \times 16$ cells, \num{4} nodes $32\times 32 \times 16$ cells and so on. Additionally, the domain length is doubled in order to double the number of particles at the same density and to keep the computational load per core constant. The number of particles within the elementary unit is \num{25} million and is of course also doubled with each doubling of the cores. The average number of communication partners for each core is always 44 and a total number of 100 time steps is carried out during the test.

\subsection{Supersonic Flow around \texorpdfstring{\ang{70}}{70} Sphere-Cone}
\label{sec:testcases:msl}
The second test case represents the near-application setup of a supersonic flow. The geometry, dimensions and the test case itself are adapted from the paper by~\citet{hollis2017blunt}. The test vehicle is a \ang{70} sphere-cone blunt-body in an high-enthalpy carbon-dioxide flow. The inflow at an angle of attack has a temperature of $T_\infty=\SI{126}{K}$, a velocity of $u_\infty=\SI{2030}{m/s}$ and a density of $\rho_\infty=\SI{5.9E-3}{kg/m^3}$ resulting in a Mach number of $M_\infty=11.4$. The flow is again simulated with the SP-BGK method.

\begin{figure}[htb!]
  \begin{subfigure}[c]{0.5\textwidth}\centering
    \includegraphics[height=7cm]{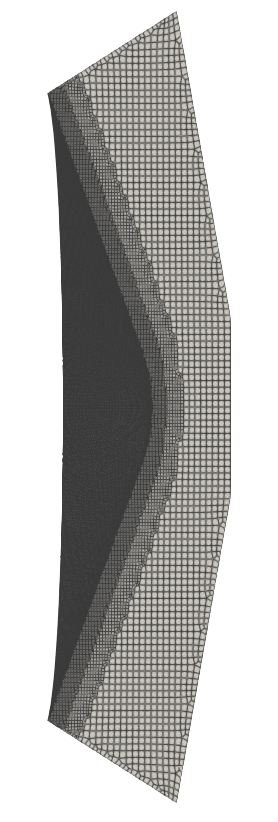}
    \caption{Side view.}
  \end{subfigure}
  \begin{subfigure}[c]{0.5\textwidth}\centering
    \includegraphics[height=7cm]{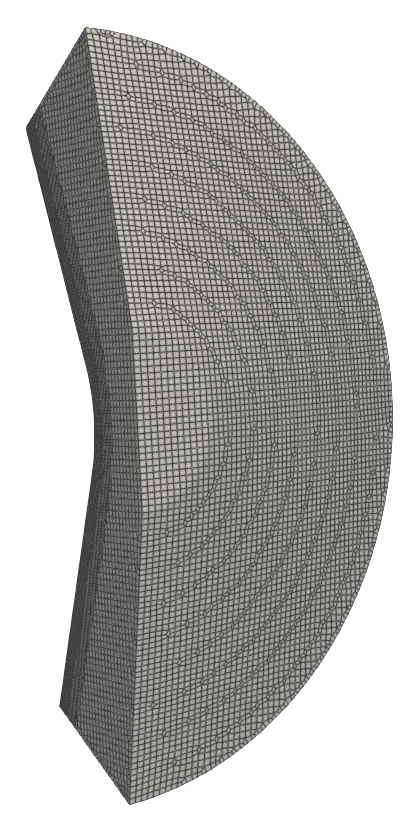}
    \caption{Front view.}
  \end{subfigure}
  \caption{Mesh of the supersonic flow test case with \num{1638637} hexahedral cells.}
  \label{fig:msl:mesh}
\end{figure}

The 3D grid with a total of \num{1638637} hexahedral cells is shown in~\cref{fig:msl:mesh}. The particle number in the simulation is \num{1.25E8}, where the time step to resolve the stiff BGK relaxation term (see~\citet{pfeiffer2018particle}) is chosen to be $\Delta t =\SI{3E-10}{s}$.
The average number of communication partners for each core ranges between \num{15} and \num{21}, increasing with the number of cores and a total number of \num{1000} time steps is carried out during the test.

\begin{figure}[htb!]
  \begin{subfigure}[c]{0.33\textwidth}\centering
    \begin{tikzpicture}[scale=0.8]
\begin{axis}[
  scaled ticks=false,
  tick label style={/pgf/number format/fixed},
  axis equal image,
  axis on top,
  xmin=-0.025,
  xmax=0.033,
  xtick={-0.02,0.03},
  ymin=-0.125,
  ymax=0.125,
  xlabel={$x$ $[\si{\metre}]$},
  ylabel={$y$ $[\si{\metre}]$},
	height=11cm,
colormap={rgb}{rgb255=(71,71,219) rgb255=(0,0,91) rgb255=(0,255,255) rgb255=(0,127,0) rgb255=(255,255,0) rgb255=(255,96,0) rgb255=(107,0,0) rgb255=(224,76,76)},
	colorbar right,
	point meta min=120,
	point meta max=2200,
	colorbar style={
      font=\small,
			scaled x ticks = false,
			height=3.5cm,
			at={(rel axis cs:1.0,0)},
			anchor=south west,
			title={$T_\mathrm{trans}$  $[\si{\kelvin}]$},
	    title style={xshift=14pt,yshift=2pt},
		}]	
	\addplot graphics [xmin=-0.025,xmax=0.033,ymin=-0.125,ymax=0.125] {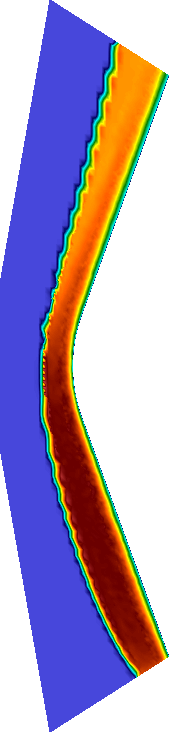};
\end{axis}
\end{tikzpicture}
    \caption{Temperature.}
  \end{subfigure}
  \begin{subfigure}[c]{0.33\textwidth}\centering
    \begin{tikzpicture}[scale=0.8]
\begin{axis}[
  scaled ticks=false,
  tick label style={/pgf/number format/fixed},
  axis equal image,
  axis on top,
  xmin=-0.025,
  xmax=0.033,
  xtick={-0.02,0.03},
  ymin=-0.125,
  ymax=0.125,
  xlabel={$x$ $[\si{\metre}]$},
  ylabel={$y$ $[\si{\metre}]$},
	height=11cm,
colormap={rgb}{rgb255=(71,71,219) rgb255=(0,0,91) rgb255=(0,255,255) rgb255=(0,127,0) rgb255=(255,255,0) rgb255=(255,96,0) rgb255=(107,0,0) rgb255=(224,76,76)},
	colorbar right,
	point meta min=6.8E22,
	point meta max=2.3E24,
	colorbar style={
      font=\small,
			height=3.5cm,
			at={(rel axis cs:1.0,0)},
			anchor=south west,
			title={$n$ $[\si{\per\cubic\metre}]$},
	    title style={xshift=14pt,yshift=2pt},
		}]	
	\addplot graphics [xmin=-0.025,xmax=0.033,ymin=-0.125,ymax=0.125] {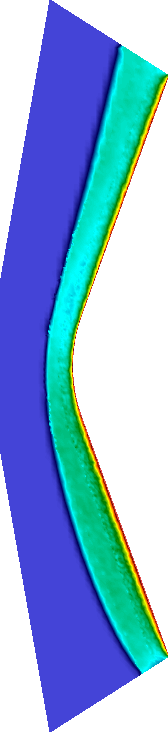};
\end{axis}
\end{tikzpicture}
    \caption{Particle Density.}
  \end{subfigure}
  \begin{subfigure}[c]{0.33\textwidth}\centering
    \begin{tikzpicture}[scale=0.8]
\begin{axis}[
  scaled ticks=false,
  tick label style={/pgf/number format/fixed},
  axis equal image,
  axis on top,
  xmin=-0.025,
  xmax=0.033,
  xtick={-0.02,0.03},
  ymin=-0.125,
  ymax=0.125,
  xlabel={$x$ $[\si{\metre}]$},
  ylabel={$y$ $[\si{\metre}]$},
	height=11cm,
colormap={rgb}{rgb255=(71,71,219) rgb255=(0,0,91) rgb255=(0,255,255) rgb255=(0,127,0) rgb255=(255,255,0) rgb255=(255,96,0) rgb255=(107,0,0) rgb255=(224,76,76)},
	colorbar right,
	point meta min=2,
	point meta max=3500,
	colorbar style={
      font=\small,
			height=3.5cm,
			at={(rel axis cs:1.0,0)},
			anchor=south west,
			title={$N_p$ $[-]$},
	    title style={xshift=14pt,yshift=2pt},
		}]	
	\addplot graphics [xmin=-0.025,xmax=0.033,ymin=-0.125,ymax=0.125] {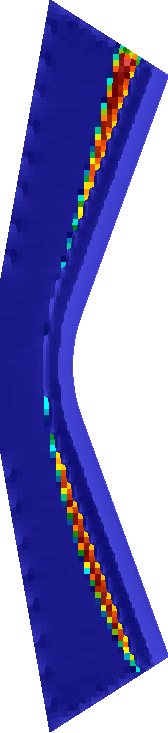};
\end{axis}
\end{tikzpicture}
    \caption{Particle Number per Cell.}
  \end{subfigure}
  \caption{Results of the flow field of the supersonic flow test case.}
  \label{fig:msl:solution}
\end{figure}

\Cref{fig:msl:solution} shows example results of temperature, particle density and particle numbers per cell $N_p$. Due to the high Mach number, a shock region is formed in front of the cone. This leads to an increase in temperature in the shock and a buildup of density and thus of particles in the post-shock region. The result is a strong imbalance of the number of particles in the grid depending on the flow and the cell size. Finally, the density within the flow varies over two orders of magnitude, thus the number of particles per cell ranges from \numrange{5}{3500}.

\subsection{140 GHz Gyrotron Resonator}
\label{sec:testcases:gyrotron}
\begin{figure}[htb!]
  \begin{subfigure}[c]{0.4\textwidth}\centering
    \includegraphics[height=5cm]{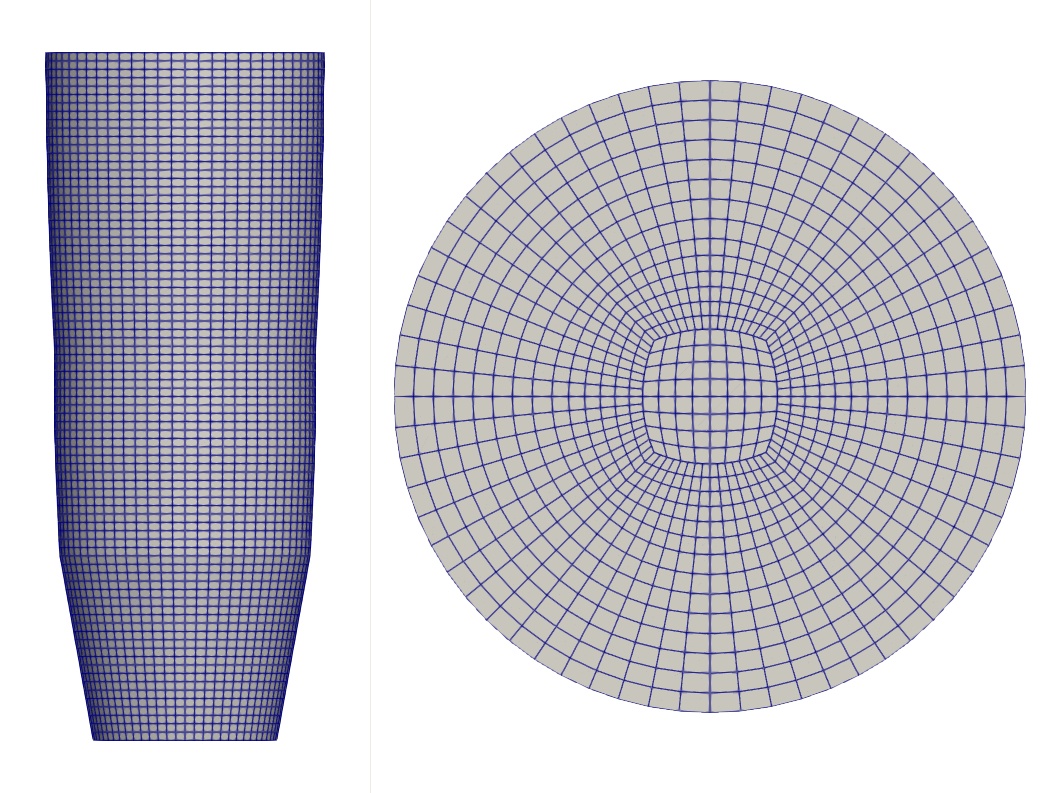}
    \caption{View from the side (left) indicating the contour of geometry and view of the exit plane (right) with the largest diameter of the cylindrical geometry.
    The exit plane corresponds to the largest cross-section of the simulation domain.}
  \end{subfigure}
  \hspace*{5mm}
  \begin{subfigure}[c]{0.5\textwidth}\centering
    \includegraphics[height=5cm]{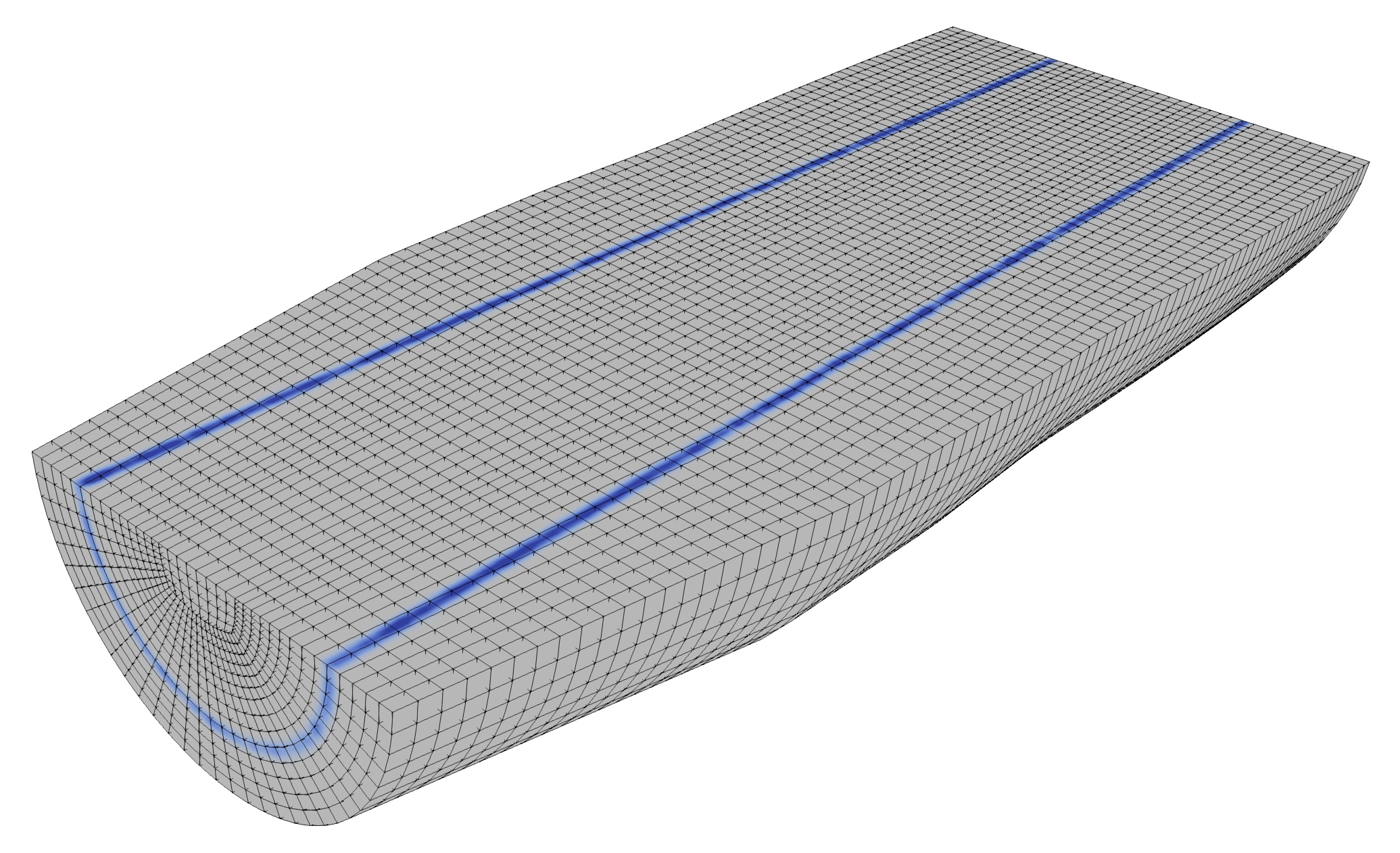}
    \caption{Clipped domain showing the charge density distribution within the domain created by the electron hollow beam.
      Elements that contain particles experience the highest workload.
    The field solver is, of course, executed for all elements in the domain.}
\label{fig:gyrotron:meshb}
  \end{subfigure}
  \caption{Mesh of the gyrotron test case with \num{78720} hexahedral cells and resulting charge density distribution indicating the location of the electron hollow beam.}
  \label{fig:gyrotron:mesh}
\end{figure}

The third test case is a gyrotron resonator operating at 140 GHz.
The details of the setup are found in~\cite{copplestone2019}, an adaption of the original setup found in~\cite{StockNeudorfer2012}.
The geometry resembles a tapered hollow cylinder, the diameter of which increases along the symmetry axis as depicted in~\cref{fig:gyrotron:mesh}. The geometry has a length of \SI{0.108035}{\metre}, a maximum diameter of \SI{0.0438134}{\metre}, and the mesh consists of \num{78720} hexahedral cells.
On the entry plane, an appropriate number of electrons is emitted in each time step to obtain a constant electron current of \SI{44}{\ampere}.
The electrons are created on a larger circle with a radius of \SI{10.1}{\milli\metre} with second smaller gyro radius of \SI{0.14168}{\milli\metre} to create a hollow electron beam.
An axial magnetic field of \SI{5.587}{\tesla} is applied, which forces the electrons to gyrate around the magnetic field lines with the specific radius given by the magnitude of the magnetic field.
Maxwell's equations are solved on the grid using a spatial order of $\mathcal{O}(5)$ and a low-storage Runge-Kutta method for time integration of order $\mathcal{O}(4)$.
The resulting charge density within the domain that is created by the electron hollow beam is shown in~\cref{fig:gyrotron:mesh}.
Only the elements that show a charge density contain simulation particles.
Therefore, the majority of the simulation domain is empty, leading to a strong workload imbalance between empty elements and those containing simulation particles.
This imbalance is addressed by the timer-based dynamic load balancing as described in~\cref{sec:theory:loadbalancing}, which assigns weights to each element and partitions the complete domain into segments such that each segment has the same computational load.
The average number of communication partners for each core ranges between \num{19} and \num{80}, increasing with the number of cores and a total number of \num{522} time steps is carried out during the test.

\section{Results}
\label{sec:results}
In the following, this paper will not interpret the results physically, but will examine the problematics of such flows for parallelization. First, the scaling of the initialization phase, i.e., the construction of the halo region and communicators, is examined. This is followed by an evaluation of the calculation phase, considering aspects such as field/particle operator and load imbalance.
The total number of elements and particles as well as the initialization and execution times (without I/O and initialization) are summarized for all test cases in~\cref{tab:results:times}.

\begin{table}[htb!]
\setlength{\multiwidth}{\widthof{\num{115000000}}}
\renewcommand{\arraystretch}{0.85}
\centering
\resizebox{\textwidth}{!}{\begin{tabular}{c|cccccccccccc}
\hline
\scriptsize{Cores}                                                    &                 & \num{128}     & \num{256}            & \num{512}             & \num{1024}            & \num{2048}            & \num{4096}      & \num{8192}       & \num{16384}  & \num{32768}  & \num{65536}  & \num{131072}  \\ \hline
\multirow{3}{*}{\rotatebox[origin=c]{90}{\scriptsize Elements}}       & adiabatic box   & \num{4096}    & \num{8192}           & \num{16384}           & \num{32768}           & \num{65536}           & \num{131072}    & \num{262144}     & \num{524288} & \num{1048576}& \num{2097152}& \num{4194304} \\
                                                                      & supersonic flow & \multicolumn{10}{c}{\textleftarrow~\makebox[\multiwidth][c]{\num{1638637}}~\textrightarrow}                                                                                                                    \\
                                                                      & gyrotron        & \multicolumn{10}{c}{\textleftarrow~\makebox[\multiwidth][c]{\num{78720}}~\textrightarrow}                                                                                                                      \\ \hline
\multirow{3}{*}{\rotatebox[origin=c]{90}{\scriptsize Particles}}      & adiabatic box   & \num{25e6}    & \num{50e6}           & \num{100e6}           & \num{200e6}           & \num{400e6}           & \num{800e6}     & \num{1600e6}     & \num{3200e6} & \num{6400e6} & \num{12800e6} & \num{25600e6}\\
                                                                      & supersonic flow & \multicolumn{10}{c}{\textleftarrow~\makebox[\multiwidth][c]{\num{125000000}}~\textrightarrow}                                                                                                                  \\
                                                                      & gyrotron        & \multicolumn{10}{c}{\textleftarrow~\makebox[\multiwidth][c]{\num{354861}}~\textrightarrow}                                                                                                                     \\ \hline
\multirow{3}{*}{\rotatebox[origin=c]{90}{\scriptsize Init. Time [s]}} & adiabatic box   & \num{1.7}     & \num{2.3}            & \num{3.6}             & \num{3.8}             & \num{4.7}             & \num{6.4}       & \num{9.6}        & \num{17.8}   & \num{41.6}   & \num{115.9}  & \num{1030.6}  \\
                                                                      & supersonic flow & \num{43.6}    & \num{45.5}           & \num{26.2}            & \num{22.7}            & \num{22.5}            & \num{23.1}      & \num{28.0}       & \num{57.0}   & -            & -            & -             \\
                                                                      & gyrotron        & \num{7.47}    & \num{11.88}          & \num{9.14}            & \num{8.10}            & \num{82.30}           & \num{8.43}      & \num{11.32}      & -            & -            & -            & -             \\ \hline
\multirow{3}{*}{\rotatebox[origin=c]{90}{\scriptsize Comp. Time [s]}} & adiabatic box   & \num{31.7}    & \num{32.3}           & \num{33.2}            & \num{33.3}            & \num{35.4}            & \num{35.4}      & \num{36.4}       & \num{40.3}   & \num{41.2}   & \num{42.1}   & \num{42.6}    \\
                                                                      & supersonic flow & \num{1743.2}  & \num{893.3}          & \num{449.8}           & \num{229.6}           & \num{115.0}           & \num{61.5}      & \num{35.6}       & \num{41.1}   & -            & -            & -             \\
                                                                      & gyrotron        & \num{1475.38} & \num{837.42}         & \num{468.44}          & \num{236.87}          & \num{132.87}          & \num{72.54}     & \num{61.77}      & -            & -            & -            & -             \\ \hline
\end{tabular}}
\caption{Number of elements/particles used in the test cases and resulting computational/initialization times.
Computational times correspond to the total elapsed wall time without I/O and initialization.
The values are averaged over multiple runs to reduce the variation in the timings.}
\label{tab:results:times}
\end{table}

\subsection{Initialization}
\label{sec:scaling:initialization}
\begin{figure}[htb]
  \begin{subfigure}[t]{.33\textwidth}\centering
    \def\X{"Procs"}
\def\MaxNodes{1024}
\def\datafile{pgfplots/BGK_Box_scaling.csv}
\def\datafileII{pgfplots/BGK_Box_InitScaling.csv}
\def\datafileV{pgfplots/BGK_Box_InitScalingOld.csv}
\def\datafileX{pgfplots/BGK_Box_InitScaling-with-mesh-read.csv}

\def\mean{average}

\begin{tikzpicture}
\pgfplotsset{table/col sep=comma}

\begin{axis}[
  xmode=log,ymode=log,
  scaled ticks=false,
  xmin=0.5,xmax=2*\MaxNodes,
  ymin=1,ymax=2000,
  width=0.9\textwidth,
  height=0.9\textwidth,
  xlabel={\# of Cores / 128},
  ylabel={Init Time [s]},
  xticklabel style = {yshift=-0.5em},
  ]; 

\addplot+[blue,mark=o*,only marks,mark size=1.5pt,mark options={fill=black}] table [x expr=\thisrow{\X}/128.0, y expr={\thisrow{"\mean"}},text special chars={\#}]{\datafileX};
\addplot+[black,mark=square*,only marks,mark size=1.5pt,mark options={fill=black}] table [x expr=\thisrow{\X}/128.0, y expr={\thisrow{"\mean"}},text special chars={\#}]{\datafileII};
\draw    [black,dotted] (32,1) -- (32,200);
\end{axis}
\end{tikzpicture}
    \caption{Adiabatic box test case.}
    \label{fig:init:adiabticbox}
  \end{subfigure}
  \begin{subfigure}[t]{.33\textwidth}\centering
    \def\X{"Procs"}
\def\MaxNodes{128}
\def\datafile{pgfplots/MSL_StrongScaling.csv}
\def\datafileII{pgfplots/MSL_InitScaling.csv}
\def\datafileIII{pgfplots/MSL_InitScaling_128.csv}

\begin{tikzpicture}[]
\pgfplotsset{table/col sep=comma}

\begin{axis}[
  xmode=log,ymode=linear,
  scaled ticks=false,
  xmin=0.5,xmax=2*\MaxNodes,
  ymin=20,ymax=70,
  width=0.9\textwidth,
  height=0.9\textwidth,
  xlabel={\# of Cores / 128},
  ylabel={Init Time [s]},
  xticklabel style = {yshift=-0.5em},
  ]

\addplot+[black,mark=square*,only marks,mark size=1.5pt,mark options={fill=black}] table [x expr=\thisrow{\X}/128.0, y expr={\thisrow{"average"}},text special chars={\#}]{\datafileII};
\addplot [only marks,mark=-,mark size=1.5pt,thick,error bars/.cd,y dir=both,y explicit,error bar style={line width=1pt,solid,mark size=1.5pt},]   table [x expr=\thisrow{\X}/128.0,y expr={\thisrow{"median"}},y error minus expr={\thisrow{"median"}-\thisrow{"min"}},y error plus expr={\thisrow{"max"}-\thisrow{"median"}},col sep=comma,] {\datafileIII};
\draw    [black,dotted] (32,.5) -- (32,2*\MaxNodes);
\end{axis}
\end{tikzpicture}
    \caption{Supersonic flow test case.}
    \label{fig:init:msl}
  \end{subfigure}
  \begin{subfigure}[t]{.33\textwidth}\centering
    \def\X{"Procs"}
\def\MaxNodes{128}
\def\datafile{pgfplots/Gyrotron_StrongScaling.csv}

\begin{tikzpicture}
\pgfplotsset{table/col sep=comma}

\begin{axis}[
  xmode=log,ymode=linear,
  scaled ticks=false,
  xmin=0.5,xmax=2*\MaxNodes,
  ymin=5,ymax=20,
  width=0.9\textwidth,
  height=0.9\textwidth,
  xlabel={\# of Cores / 128},
  ylabel={Init Time [s]},
  xticklabel style = {yshift=-0.5em},
  ytick={5,10,15,20}
  ]; 

\addplot+[black,mark=square*,only marks,mark size=1.5pt,mark options={fill=black}] table [x expr=\thisrow{\X}/128.0, y expr={\thisrow{"Init Start"}},text special chars={\#}]{\datafile};
\draw    [black,dotted] (32,5) -- (32,20);
\end{axis}
\end{tikzpicture}
    \caption{Gyrotron test case.}
    \label{fig:init:gyrotron}
  \end{subfigure}
  \caption{Initialization times of the weak scaling of the adiabatic box (left) test case and the strong scaling of the supersonic flow (middle) and the gyrotron resonator (right) test cases.}
  \label{fig:init}
\end{figure}
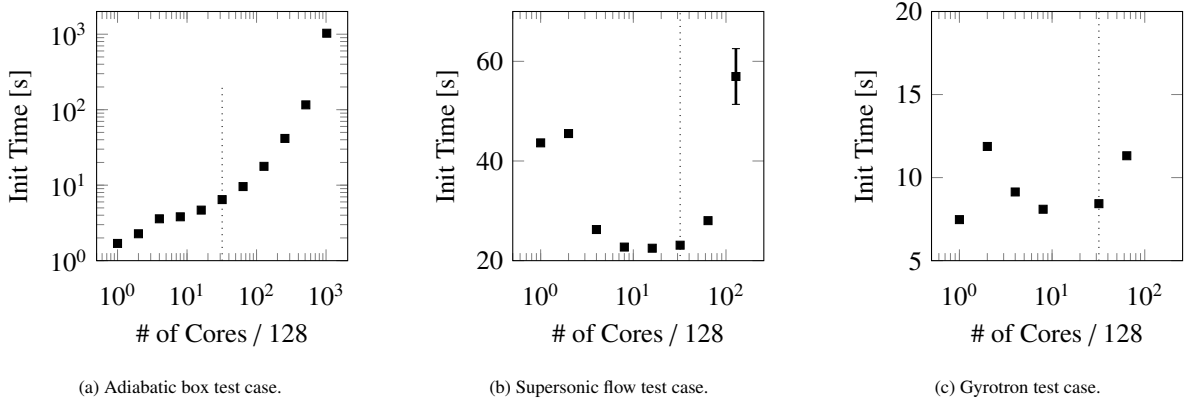

Within the context of this paper, the initialization comprises the complete code startup including the initial insertion of the particles or the corresponding restart routines in the case of a continued simulation. The initialization time for different numbers of processors for all test cases is depicted in~\cref{fig:init}. In accordance with the parallelization concept outlined in~\cref{sec:implementation:halo}, a constant number of total mesh elements should result in an approximately constant time for the first step of the halo element search algorithm,  regardless of the number of compute nodes used. Time spent in subsequent initialization routines will decrease as more elements reside outside the halo region and can be discarded, given a sufficiently small halo distance compared to the overall grid dimensions. At the same time, a less than linear increase of initialization time is expected for an increasing number of total mesh elements, as only the first step of the halo element search algorithm scales with the amount of grid cells, provided the retained number of local and halo elements remains the same.

\paragraph{Weak Scaling}
Results for the weak scaling test case of the adiabatic box are shown in~\cref{fig:init:adiabticbox}. As the effort to the initial mapping from mesh elements to the corresponding BGM cells grows while the remaining routines scale out, the overall time required increases at a slope with gradient $< 1$ until the negative effects of decreasing interconnect bandwidth become dominant.

\paragraph{Strong Scaling}
As expected, the initialization time in the strong scaling for the supersonic flow in \cref{fig:init:msl} and the gyrotron in \cref{fig:init:gyrotron} remains in a similar range, irrespective of the number of compute nodes. Initially, the initialization time decreases as the number of processors increases, since the communication-free aspects of the initialization process can be effectively distributed. However, with a greater number of processors, the communication time of the remaining initialization routines becomes more significant and the initialization time rises again, nonetheless remaining within the same order of magnitude.

One exception is the run on a single node for both setups. For this case the initialization time of the gyrotron is shorter than with two nodes. The advantageous effect is that in the case of a single node, all grid cells are automatically compute node local elements and thus no halo region needs to be constructed, thus saving initialization time. Another point is visible for the supersonic test case. As the node count exceeds \num{16}, the peculiarity of the test system becomes apparent. Since \num{16} nodes are directly connected to a switch, each doubling of the node numbers introduces an additional hop, leading to an increase in the initialization time as communication calls now have reduced bandwidth available. This yields an increased variation in initialization and calculation times and therefore, the minimum and maximum as well as the average values are depicted.

\subsection{Simulation Performance}
\paragraph{Weak Scaling}
The parallel efficiency $\eta_N$ for the weak scaling was determined by
\begin{equation}
  \eta_N=\frac{t_{128}}{t_N},
\end{equation}
where $t_{128}$ and $t_N$ are the computational time using \num{128} and $N$ cores, respectively. Since the problem size increases linearly with the number of cores used, the parallel efficiency should ideally remain around one.

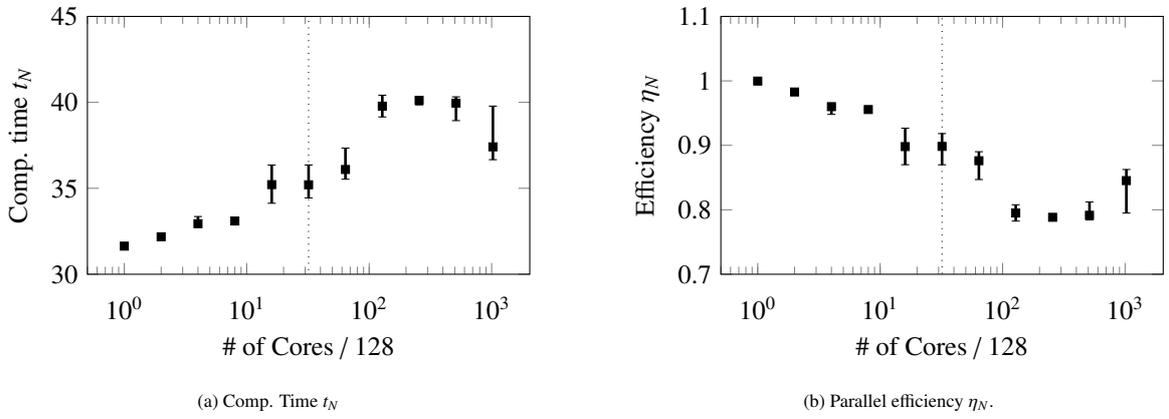
\begin{figure}[htb!]
  \begin{subfigure}[t]{0.5\textwidth}\centering
    \def\X{"Procs"}
\def\RunTime{"Runtime"}
\def\IniTime{"Init Start"}
\def\baselineRunTime{31.25}
\def\baselineIniTime{2}
\def\MaxProcs{1024}
\def\datafile{pgfplots/BGK_Box_scaling.csv}

\def\datafileII{pgfplots/BGK_Box_calctime.csv}
\def\datafileIII{pgfplots/BGK_Box_calctime_64.csv}

\def\datafileXII{pgfplots/BGK_Box_PID.csv}
\def\datafileXIII{pgfplots/BGK_Box_PID_64.csv}

\def\mean{average}

\begin{tikzpicture}
\pgfplotsset{table/col sep=comma}

\begin{axis}[
  xmode=log,ymode=linear,
  scaled ticks=false,
  xmin=.5,xmax=2*\MaxProcs,
  ymin=30,ymax=45,
  width=0.9\textwidth,
  height=5cm,
  xlabel={\# of Cores / 128},
  ylabel={Comp. time $t_N$},
  xticklabel style = {yshift=-0.5em},
  ]; 
\draw    [black,dotted] (32,1) -- (32,200);
\addplot+[black,mark=square*,only marks,mark size=1.5pt,mark options={fill=black}] table [x expr=\thisrow{\X}/128.0, y expr={\thisrow{"\mean"}*25600},text special chars={\#}]{\datafileXII}; 
\addplot [only marks,mark=-,mark size=1.5pt,thick,error bars/.cd,y dir=both,y explicit,error bar style={line width=1pt,solid,mark size=1.5pt},] table [x expr=\thisrow{\X}/128.0,
y expr={\thisrow{"\mean"}*25600},
y error minus expr={(\thisrow{"\mean"}-\thisrow{"min"})*25600},
y error plus expr={(\thisrow{"max"}-\thisrow{"\mean"})*25600},col sep=comma,] {\datafileXII};

\end{axis}
\end{tikzpicture}
    \caption{Comp. Time $t_N$}
  \end{subfigure}
  \begin{subfigure}[t]{0.5\textwidth}\centering
    \def\X{"Procs"}
\def\RunTime{"Runtime"}
\def\IniTime{"Init Start"}
\def\baselineRunTime{35.50}
\def\baselineIniTime{2.2}
\def\baselineV{1.284849999999999924e-03} 
\def\baseline{1.235360000000000017e-03} 
\def\MaxProcs{1024}
\def\datafile{pgfplots/BGK_Box_scaling.csv}
\def\datafileII{pgfplots/BGK_Box_PID.csv}
\def\datafileIII{pgfplots/BGK_Box_PID_64.csv}

\def\mean{average}

\begin{tikzpicture}
\pgfplotsset{table/col sep=comma}

\begin{axis}[
  xmode=log,ymode=linear,
  scaled ticks=false,
  xmin=.5,xmax=2*\MaxProcs,
  ymin=0.7,ymax=1.1,
  width=0.9\textwidth,
  height=5cm,
  xlabel={\# of Cores / 128},
  ylabel={Efficiency $\eta_N$},
  xticklabel style = {yshift=-0.5em},
  ]; 

\draw    [black,dotted] (32,0) -- (32,1.1);
\addplot+[black,mark=square*,only marks,mark size=1.5pt,mark options={fill=black}] table [x expr=\thisrow{\X}/128.0, y expr={\baseline/\thisrow{"\mean"}},text special chars={\#}]{\datafileII};
\addplot [only marks,mark=-,mark size=1.5pt,thick,error bars/.cd,y dir=both,y explicit,error bar style={line width=1pt,solid,mark size=1.5pt},]  table [x expr=\thisrow{\X}/128.0,
y expr={\baseline/\thisrow{"\mean"}},
y error minus expr={\baseline/\thisrow{"\mean"}-\baseline/\thisrow{"min"}},
y error plus expr={\baseline/\thisrow{"max"}-\baseline/\thisrow{"\mean"}},col sep=comma,] {\datafileII};
\end{axis}
\end{tikzpicture}
    \caption{Parallel efficiency $\eta_N$.}
  \end{subfigure}
  \caption{Weak scaling of the adiabatic box test case from computational time $t_N$, which is calculated from the total run time of the application minus the initialization time.}
  \label{fig:speedup:adbox}
\end{figure}

Already with low core counts, weak scaling of the adiabatic box in~\cref{fig:speedup:adbox} shows a decreasing efficiency with an increasing number of cores. As the setup is inherently ideally load balanced, the increase in computing time is assumed to be related to an increasing time for particle communication via the nodes. However, the parallel efficiency with \num{32} nodes (\num{4096} cores) is still around \num{0.9}, which is satisfactory. Reaching \num{64} nodes, the influence of the reduced bandwidth becomes visible due to the higher particle density in the domain. For even higher node numbers, the efficiency increases as the weak scaling results benefit from communication locality, thereby compensating for the interconnect penalty.


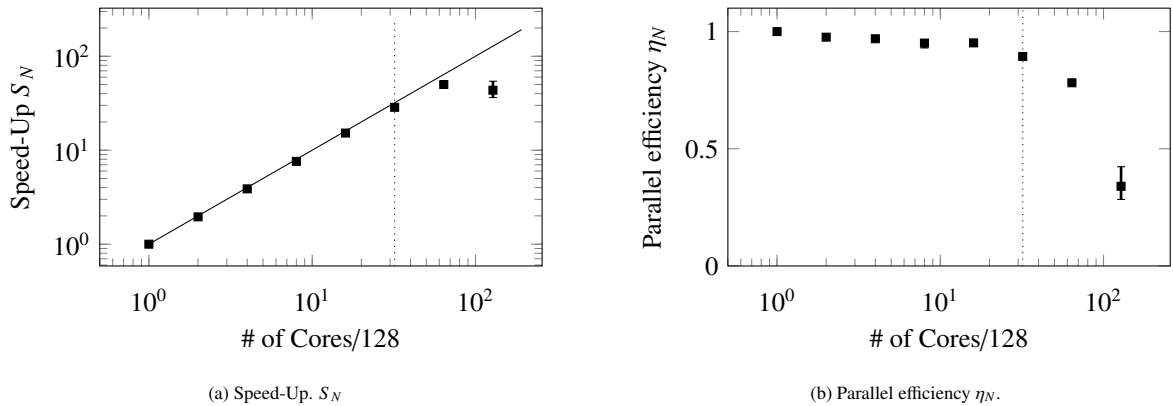
\begin{figure}[htb!]
  \begin{subfigure}[t]{0.5\textwidth}\centering
    \def\X{"Procs"}
\def\baseline{1.80541666666667E-05}
\def\baseline{0.0000180541666666667} 
\def\MaxNodes{128}
\def\datafile{pgfplots/MSL_StrongScaling.csv}
\def\datafileII{pgfplots/MSL_PID.csv}

\def\mean{average}

\begin{tikzpicture}[]
\pgfplotsset{table/col sep=comma}

\begin{axis}[
  xmode=log,ymode=log,
  scaled ticks=false,
  xmin=.5,xmax=2*\MaxNodes,
  width=0.9\textwidth,
  height=5cm,
  xlabel={\# of Cores/128},
  ylabel={Speed-Up $S_N$},
  xticklabel style = {yshift=-0.5em},
]

\draw    [black,dotted] (32,.5) -- (32,2*\MaxNodes);
\addplot+[black,mark=square*,only marks,mark size=1.5pt,mark options={fill=black}] table [x expr=\thisrow{\X}/128.0, y expr={\thisrow{\X}*\baseline/\thisrow{"\mean"}/128.0},text special chars={\#}]{\datafileII};
\addplot [only marks,mark=-,mark size=1.5pt,thick,error bars/.cd,y dir=both,y explicit,error bar style={line width=1pt,solid,mark size=1.5pt},]
      table [x expr=\thisrow{\X}/128.0,
      y expr={\thisrow{\X}*\baseline/\thisrow{"\mean"}/128.0},
      y error minus expr={\thisrow{\X}*\baseline/\thisrow{"\mean"}/128.0-\thisrow{\X}*\baseline/\thisrow{"min"}/128.0},
      y error plus expr={\thisrow{\X}*\baseline/\thisrow{"max"}/128.0-\thisrow{\X}*\baseline/\thisrow{"\mean"}/128.0},col sep=comma,] {\datafileII};

\addplot[domain=1:192]{x};
\end{axis}
\end{tikzpicture}
    \caption{Speed-Up. $S_N$}
    \label{fig:speedup:msl:strong}
  \end{subfigure}
  \begin{subfigure}[t]{0.5\textwidth}\centering
    \def\X{"Procs"}
\def\baseline{1.80541666666667E-05}
\def\baseline{0.0000180541666666667} 
\def\MaxNodes{128}
\def\datafile{pgfplots/MSL_StrongScaling.csv}
\def\datafileII{pgfplots/MSL_PID.csv}
\def\datafileIII{pgfplots/MSL_PID.csv}

\def\mean{average}

\begin{tikzpicture}[]
\pgfplotsset{table/col sep=comma}

\begin{axis}[
  xmode=log,ymode=linear,
  xmin=.5,xmax=2*\MaxNodes,
  ymin=0.0,ymax=1.1,
  width=0.9\textwidth,
  height=5cm,
  xlabel={\# of Cores/128},
  ylabel={Parallel efficiency $\eta_N$},
  xticklabel style = {yshift=-0.5em},
]
\addplot+[black,mark=square*,only marks,mark size=1.5pt,mark options={fill=black}] table [x expr=\thisrow{\X}/128.0, y expr={\baseline/\thisrow{"\mean"}},text special chars={\#}]{\datafileII};
\draw    [black,dotted] (32,.0) -- (32,1.1);
\addplot [only marks,mark=-,mark size=1.5pt,thick,error bars/.cd,y dir=both,y explicit,error bar style={line width=1pt,solid,mark size=1.5pt},]  table [x expr=\thisrow{\X}/128.0,
y expr={\baseline/\thisrow{"\mean"}},
y error minus expr={\baseline/\thisrow{"\mean"}-\baseline/\thisrow{"min"}},
y error plus expr={\baseline/\thisrow{"max"}-\baseline/\thisrow{"\mean"}},col sep=comma,] {\datafileII};
\end{axis}
\end{tikzpicture}
    \caption{Parallel efficiency $\eta_N$.}
    \label{fig:speedup:msl:eff}
  \end{subfigure}
  \caption{Strong scaling of the supersonic flow test case.}
  \label{fig:speedup:msl}
\end{figure}

\begin{figure}[htb!]
  \begin{subfigure}[t]{0.5\textwidth}\centering
    \def\X{"Procs"}
\def\baseline{0.0000367625}
\def\MaxProcs{128}
\def\datafile{pgfplots/Gyrotron_StrongScaling.csv}
\def\datafileII{pgfplots/Gyrotron_PID.csv}

\def\mean{average}

\begin{tikzpicture}
\pgfplotsset{table/col sep=comma}

\begin{axis}[
  xmode=log,ymode=log,
  scaled ticks=false,
  xmin=.5,xmax=2*\MaxProcs,
  width=0.9\textwidth,
  height=5cm,
  xlabel={\# of Cores/128},
  ylabel={Speed-Up $S_N$},
  xticklabel style = {yshift=-0.5em},
]

\draw    [black,dotted] (32,.5) -- (32,2*\MaxProcs);
\addplot+[black,mark=square*,only marks,mark size=1.5pt,mark options={fill=black}] table [x expr=\thisrow{\X}/128.0, y expr={\thisrow{\X}*\baseline/\thisrow{"\mean"}/128.0},text special chars={\#}]{\datafileII};
\addplot [only marks,mark=-,mark size=1.5pt,thick,error bars/.cd,y dir=both,y explicit,error bar style={line width=1pt,solid,mark size=1.5pt},]
      table [x expr=\thisrow{\X}/128.0,
      y expr={\thisrow{\X}*\baseline/\thisrow{"\mean"}/128.0},
      y error minus expr={\thisrow{\X}*\baseline/\thisrow{"\mean"}/128.0-\thisrow{\X}*\baseline/\thisrow{"min"}/128.0},
      y error plus expr={\thisrow{\X}*\baseline/\thisrow{"max"}/128.0-\thisrow{\X}*\baseline/\thisrow{"\mean"}/128.0},col sep=comma,] {\datafileII};

\addplot[domain=1:192]{x};
\end{axis}
\end{tikzpicture}
    \caption{Speed-Up. $S_N$}
    \label{fig:speedup:gyrotron:strong}
  \end{subfigure}
  \begin{subfigure}[t]{0.5\textwidth}\centering
    \def\X{"Procs"}
\def\RunTime{"Runtime"}
\def\IniTime{"Init Start"}
\def\baseline{0.0000367625}
\def\MaxProcs{128}
\def\datafile{pgfplots/Gyrotron_StrongScaling.csv}
\def\datafileII{pgfplots/Gyrotron_PID.csv}

\def\mean{average}

\begin{tikzpicture}
\pgfplotsset{table/col sep=comma}

\begin{axis}[
  xmode=log,ymode=linear,
  scaled ticks=false,
  xmin=.5,xmax=2*\MaxProcs,
  ymin=0,ymax=1.1,
  width=0.9\textwidth,
  height=5cm,
  xlabel={\# of Cores / 128},
  ylabel={Efficiency $\eta_N$},
  xticklabel style = {yshift=-0.5em},
  ]; 

\draw    [black,dotted] (32,0) -- (32,1.1);

\addplot+[black,mark=square*,only marks,mark size=1.5pt,mark options={fill=black}] table [x expr=\thisrow{\X}/128.0, y expr={\baseline/\thisrow{"\mean"}},text special chars={\#}]{\datafileII};
\addplot [only marks,mark=-,mark size=1.5pt,thick,error bars/.cd,y dir=both,y explicit,error bar style={line width=1pt,solid,mark size=1.5pt},]  table [x expr=\thisrow{\X}/128.0,
y expr={\baseline/\thisrow{"\mean"}},
y error minus expr={\baseline/\thisrow{"\mean"}-\baseline/\thisrow{"min"}},
y error plus expr={\baseline/\thisrow{"max"}-\baseline/\thisrow{"\mean"}},col sep=comma,] {\datafileII};
\end{axis}
\end{tikzpicture}
    \caption{Parallel efficiency $\eta_N$.}
    \label{fig:speedup:gyrotron:eff}
  \end{subfigure}
  \caption{Strong scaling of the gyrotron test case.}
  \label{fig:speedup:gyrotron}
\end{figure}
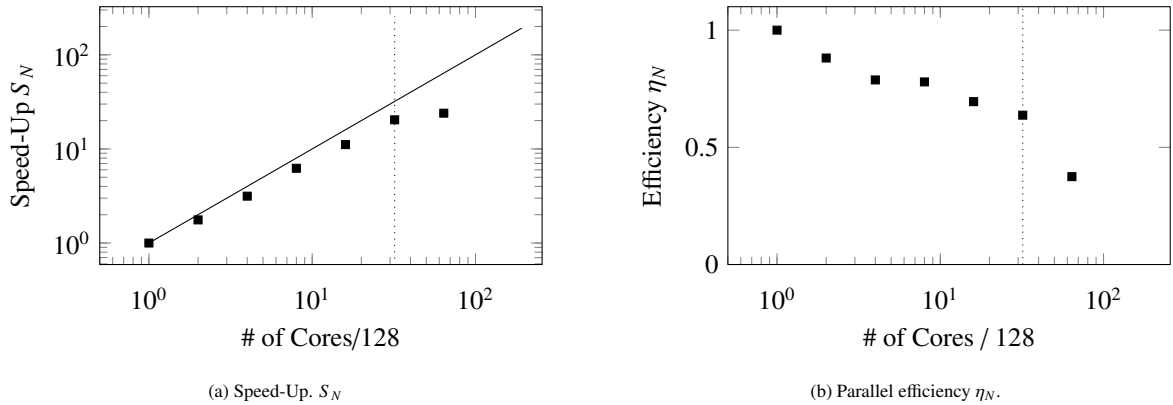

\paragraph{Strong Scaling}
The strong scaling was calculated using the basis of one node corresponding to \num{128} cores by
\begin{equation}
  S_N=\frac{t_{128}}{t_N},
\end{equation}
with the respective parallel efficiency $\eta_N$ determined by
\begin{equation}
  \eta_N=\frac{128\cdot t_{128}}{N \cdot t_N},
\end{equation}
where $t_{128}$ and $t_N$ are the computational time using \num{128} and $N$ cores, respectively. The strong scaling results of the supersonic flow and gyrotron test case are depicted in \cref{fig:speedup:msl:strong,fig:speedup:gyrotron:strong} with the parallel efficiency shown in \cref{fig:speedup:msl:eff,fig:speedup:gyrotron:eff}, respectively.

Due to the load balancing as described in~\cref{sec:theory:loadbalancing}, whereby the grid cells are weighted by the number of particles they currently contain when they are divided among the processors and the latency hiding as described in~\cref{sec:LH_BGK}, the parallel efficiency of the supersonic flow test case remains around one up to \num{16} nodes, which corresponds to \num{2048} processors. For larger processor numbers, the parallel efficiency then drops due to the large discrepancies in computational load between different regions of the computational domain. Some elements contain so many particles that they can no longer be better distributed by the load balancing and thus also the latency hiding can no longer work properly. In this case, it would be necessary to increase the number of mesh elements in the regions of high particle density, thereby allowing the computational load to be better distributed. Nevertheless, even with \num{64} nodes, which corresponds to \num{8192} processors, a parallel efficiency of \num{0.77} is still achieved for this case with a very nonuniform load distribution. This setup thus retains the better scaling properties than the adiabatic box, presumably because the lower particle number poses less requirements on the interconnect bandwidth. Nonetheless, it is still sensitive to an even further increase in latency as is seen when \num{64} nodes are exceeded.

Compared to the strong scaling of the supersonic flow case, the scaling behavior is distinctively worse in the gyrotron case. As visible in~\cref{fig:speedup:gyrotron:eff}, the parallel efficiency drops significantly faster as the number of processors increases, so that with \num{64} nodes (\num{8192} cores), a parallel efficiency of only \num{0.37} is achieved. The main reason for this is most likely the more complex load balancing for this case. On the one hand, the presence of the Maxwell solver introduces differing loads per cell between field solver and the particle solver. On the other hand, the number of particles per element differs by orders of magnitude, as shown in~\cref{fig:gyrotron:meshb}. Numerous cells contain no particles at all while only very few carry the particle beam shaped as a hollow cylinder. As a consequence, not only is an ideal load distribution increasingly difficult to achieve as the number of processors increases, but efficient latency hiding also becomes progressively unattainable. At any given time, there exists a considerable fraction of processors that partake in the particle communication without ever receiving particles in their corresponding cells. These processors thus stall the particle communication of the remaining cores, a phenomenon which cannot be hidden. Ultimately, in this test case, more and more cores have to wait for communication, resulting in a scaling which is not optimal.

\section{Conclusions}
\label{sec:conclusions}
As computers become increasingly parallel, code areas with previously negligible performance impact such as initialization become increasingly more relevant due to their influence on load balancing. In the work presented here, our aim was to contribute to this challenge by presenting a massively parallel, communication-free approach to build the halo region required for Euler-Lagrange simulations. The use of the MPI-3 shared memory model enabled us to utilize the ever increasing core count per socket without introducing additional interconnect load. Based on this programming model, we developed new methods for emission, deposition and latency hiding which were implemented in the open-source plasma dynamics framework PICLas. This framework was applied to a generic test setup as well as two practical application cases. In all setups, we were able to show respectable initialization times as long as no interconnect congestion occurred in other parts of the startup phase. Furthermore, we were able to retain good efficiency for both the weak and the strong scaling case of the BGK setup while also highlighting challenges inherent to the Euler-Lagrange setup including strong variations in particle density in the PIC setup.

In the future, we will extend our research to communication-minimizing decomposition approaches for the Euler-Lagrange codes. These pose additional challenges since the continuous and the disperse phase entail separate communication regions with machine-dependent costs, resulting in a multi-point optimization problem.

\section*{Acknowledgment} 
The authors gratefully acknowledge the support and the computing time on \hawk\ provided by the HLRS through the projects "hpcdg" and "impd".
We acknowledge PRACE for awarding us access to \hawk\ at GCS@HLRS, Germany.
Marcel Pfeiffer has received funding from the European Research Council (ERC) under the European Union’s Horizon 2020 research and innovation programme (grant agreement No. 899981 MEDUSA).

\bibliographystyle{elsarticle-num-names}
\bibliography{bibliography}

\end{document}